\documentclass[sn-aps,Numbered,pdflatex,iicol]{sn-jnl}

\usepackage{graphicx}%
\usepackage{multirow}%
\usepackage{amsmath,amssymb,amsfonts}%
\usepackage{amsthm}%
\usepackage{mathrsfs}%
\usepackage[title]{appendix}%
\usepackage{xcolor}%
\usepackage{textcomp}%
\usepackage{manyfoot}%
\usepackage{booktabs}%
\usepackage{algorithm}%
\usepackage{algorithmicx}%
\usepackage{algpseudocode}
\usepackage{listings}%
\usepackage{widetext}%
\usepackage{tikz}%
\usepackage[compat=1.1.0]{tikz-feynman}%
\usepackage{widetext}
\usepackage{multibib}

\newcommand{\tlambda}{\tilde{\lambda}}

\newcommand{\iu}{\text{i}}
\newcommand{\eu}{\text{e}}

\newcommand{\dder}[2]{\frac{\delta #1}{\delta #2}}
\newcommand{\vecc}[1]{\boldsymbol{#1}}

\newcommand{\di}{\text{d}}

\newcommand{\hphi}{{\tilde{\phi}}}
\newcommand{\vphi}{{\vecc{\phi}}}
\newcommand{\hvphi}{{\tilde{\vecc{\phi}}}}

\newcommand{\vx}{\vecc{x}}

\newcommand{\E}[1]{\mathbb{E}\left[#1\right]}

\newcommand{\tD}{\tilde{D}}
\newcommand{\Ham}{\mathcal{F}}
\newcommand{\bol}[1]{\boldsymbol{#1}}

\newcommand{\bra}[1]{\langle{#1}|}
\newcommand{\ket}[1]{|{#1}\rangle}
\newcommand{\abs}[1]{\left| {#1}\right|}
\newcommand{\braket}[2]{\langle{#1}|{#2}\rangle}

\begin{document}

\title[Article Title]{Active Ising Models of Flocking: A Field-Theoretic Approach}

\author[1,2,3]{\fnm{Mattia} \sur{Scandolo}}

\author[4]{\fnm{Johannes} \sur{Pausch}}

\author[3]{\fnm{Michael E.} \sur{Cates}}

\affil[1]{\orgdiv{Dip. di Fisica}, \orgname{Università Sapienza}, \orgaddress{\city{Rome}, \postcode{00185}, \country{Italy}}}

\affil[2]{\orgdiv{Istituto dei Sistemi Complessi}, \orgname{Consiglio Nazionale delle Ricerche}, \orgaddress{\city{Rome}, \postcode{00185}, \country{Italy}}}

\affil[3]{\orgdiv{DAMTP}, \orgname{Centre for Mathematical Sciences, University of Cambridge}, \orgaddress{\street{Wilberforce Road}, \city{Cambridge} \postcode{CB3 0WA}, \country{United Kingdom}}}

\affil[4]{\orgdiv{Department of Mathematics}, \orgname{Imperial College London}, \orgaddress{ \city{London} \postcode{SW7 2AZ}, \country{United Kingdom}}}

\abstract{Using an approach based on Doi-Peliti field theory, we study several different Active Ising Models (AIMs), in each of which collective motion (flocking) of self-propelled particles arises from the spontaneous breaking of a discrete symmetry. We test the predictive power of our field theories by deriving the hydrodynamic equations for the different microscopic choices of aligning processes that define our various models. At deterministic level, the resulting equations largely confirm known results, but our approach has the advantage of allowing systematic generalization to include noise terms. Study of the resulting hydrodynamics allows us to confirm that the various AIMs share the same phenomenology of a first order transition from isotropic to flocked states whenever the self propulsion speed is nonzero, with an important exception for the case where particles align only pairwise locally. Remarkably, this variant fails entirely to give flocking -- an outcome that was foreseen in previous work, but is confirmed here and explained in terms of the scalings of various terms in the hydrodynamic limit. Finally, we discuss our AIMs in the limit of zero self-propulsion where the ordering transition is continuous. In this limit, each model is still out of equilibrium because the dynamical rules continue to break detailed balance, yet it has been argued that an equilibrium universality class (Model C) prevails. We study field-theoretically the connection between our AIMs and Model C, arguing that these particular models (though not AIMs in general) lie outside the Model C class. We link this to the fact that in our AIMs without self propulsion, detailed balance is not merely still broken, but replaced by a different dynamical symmetry in which the dynamics of the particle density is independent of the spin state.}

\keywords{Active Ising Model, Flocking, Field Theory, Phase Transition}

\maketitle

\section{Introduction}\label{sec1}
Flocking, in which a group of self-propelled particles align and move in the same direction, is displayed by a wide variety of biological and soft-matter systems \cite{tonertu2005hydro}.  
The alignment effect creates many similarities between flocking and ferromagnetism, but flocking exhibits a richer phenomenology. Indeed, because the alignment interaction is among particle velocities rather than spatially fixed spin variables, flocking system are inherently active, and driven far from thermal equilibrium. Much recent research has addressed collective behaviour and phase transitions in this and other active matter systems, with many open questions remaining \cite{marchetti2013hydro}.
In this paper we address some questions concerning minimal models of active matter (specifically, flocking), exploiting similarities to magnetism (specifically, the Ising model). In doing so we follow a path exemplified decades ago by Fyl Pincus, who was among the first to properly explore the similarities between solid-state magnetism and liquid-crystal ordering in soft matter systems (e.g. \cite{pincus1966,pincus1969}).

In their seminal paper on flocking -- also known as collective motion -- Vicsek {\it et al.} \cite{vicsek1995novel} studied a system of active particles with fixed speed that align their velocities through a ferromagnetic interaction. The Vicsek model is thus an active version of the XY or Heisenberg model, in $2d$ and $3d$ respectively. Crucially, activity breaks the precepts of the Mermin-Wagner theorem, stabilizing the ordered phase even in two dimensions \cite{tonertu1995}. Moreover, the phase transition from disorder to the ordered (flocked) phase exhibits clear evidence of being first-order \cite{gregoire2004,chate2008collective,dicarlo2022evidence}, in contrast to the usual second-order nature of the ordering transition in equilibrium ferromagnets.

To help understand the transition, Solon and Tailleur introduced the Active Ising Model \cite{solon2013revisiting,solon2015flocking}. Here flocking can emerge only along one privileged axis (the $x$-axis, say), just as the  magnetization axis is pre-determined in the equilibrium Ising model. Although {\em the} Active Ising Model was introduced with a specific choice of spin-alignment dynamics, below we will refer to any model in which velocities locally tend to align along a fixed axis as {\em an} Active Ising Model (or AIM).

In the present work, we study the behaviour of  various AIMs through a field-theoretical approach. As in equilibrium, a main advantage of using field theory is that any large scale, collective, behaviours that do not depend on the specific microscopic details can be simpler to study and understand. This happens because, once the emergent (hydrodynamic) variables are identified, an expansion in small, slowly variating fluctuations is typically possible -- either directly, or by re-expressing as an expansion in dimensionality via the Renormalisation Group (RG).

Below, starting from the Master Equation for AIM systems, we derive a field theory through a coherent-state path-integral representation. This approach, known as Doi-Peliti field theory, offers an exact mapping between the coefficients of the microscopic model and the bare couplings of a field-theoretic action. It not only enables standard field-theoretical approximations including RG, but also gives the exact deterministic hydrodynamic equations, together with their lowest order fluctuation corrections \cite{tauber2014critical}.

Using our field theory, we will present several new results, some  in line with prior expectations based on less formal analyses, but some others contradicting such expectations. Our results add significantly to what is known about Active Ising Models, although many questions lie beyond our scope and must remain unanswered here.

The rest of this paper is structured as follows: in Sec~\ref{aimdef} we define the various Active Ising Models studied, and review their main phenomenology. In Sec~\ref{ft} we briefly review the derivation of the Doi-Peliti field theory, connecting it to physical variables through the so-called Cole-Hopf transformation. In Sec~\ref{hydro} we derive the hydrodynamic equations, together with lowest order noise terms, and give a linear stability analysis of homogeneous states. In Sec~\ref{twobody} we focus on a specific AIM in which spins align only pairwise, finding this unable to sustain flocking at any finite noise level, in agreement with previous arguments \cite{chatterjee2019threebody}. In Sec~\ref{trans} we address the continuous ordering transition of our Active Ising Models that arises in the limit of zero self-propulsion. As previously noted \cite{solon2013revisiting,solon2015flocking}, such models remain active (i.e., out of equilibrium) because the remaining combination of unbiased spin hopping and alignment already breaks detailed balance. We discuss whether this transition shares a universality class with equilibrium models as previously argued \cite{solon2015flocking} (a result also seen in other active models of Ising symmetry \cite{caballero2020stealth}). 
We establish a connection between our stochastic hydrodynamic equations and Model C (which describes an equilibrium Ising dynamics coupled to a conserved scalar density) but argue that AIMs may nonetheless inhabit a new, nonequilibrium universality class -- a view supported by explicit RG calculations that we will publish elsewhere.
Finally, in Sec~\ref{conc} we offer some concluding remarks. 

\section{Active Ising Models}\label{aimdef}
An Active Ising Model (AIM) is a minimal description of a system in which individuals align their directions of motion. Contrary to the Vicsek Model \cite{vicsek1995novel}, where collective motion may occur in any possible direction in space, in an AIM, individuals prefer to move parallel to a given axis, which we identify {\em wlog} as the $x$ axis. The {\it state} of each particle is thus defined by its lattice position and a spin variable $\pm1$ that tells which direction $\pm\hat x$ it prefers to move in. The particles reside on a $d$-dimensional square lattice {\em without any occupation number constraint} and move through space by hopping onto neighbouring sites. In the $x$ direction (only) the hopping rates are actively biased: particles with positive (negative) spin will hop preferentially towards more positive (negative) $x$ values. In all directions other than $x$, particles undergo unbiased, diffusive hopping. Finally, imitative behaviour among individuals, effectively encoded in a ferromagnetic spin alignment interaction among particles on the same site, creates a tendency towards mutual alignment and hence collective motion.

Thus an AIM represents a minimal, Ising-like model of flocking, with a discrete symmetry replacing the full rotational symmetry of the Vicsek Model. Two crucial differences between an AIM and the equilibrium Ising Model must be borne in mind: (i) an AIM has no occupancy constraint on each lattice site, and (ii) the alignment interaction occurs only between same-site particles instead of between particles on nearest neighbour sites. The former means that particles are never blocked from hopping by excluded volume, allowing a simpler treatment of the bias. The latter choice is likewise made for simplicity in the hope that same-site interactions are sufficient to describe emergent properties; in most cases one expects diffusion to mix particles enough that on-site and nearest neighbour alignment interactions are equivalent.

The state of the $k$-th particle is defined by its position on the lattice $\bol{i}^{(k)}=\left(i_1^{(k)},\dots,i_d^{(k)}\right)$ and its spin $s_k = \pm 1$. The state of the whole system can then be identified via the number of $+1$ and $-1$ spin particles on each site $\bol{i}$, respectively $n_{\bol{i}}^+$ and $n_{\bol{i}}^-$, or equivalently via the local density $\rho_{\bol{i}}=n_{\bol{i}}^++n_{\bol{i}}^-$, and magnetisation $m_{\bol{i}}=n_{\bol{i}}^+-n_{\bol{i}}^-$. With no occupational constraint, $\rho_{\bol{i}}$ has no upper bound, but the magnetisation $m_{\bol{i}}$ is  bounded by $\rho_{\bol{i}}$, since $-\rho_{\bol{i}}\leq m_{\bol{i}}\leq\rho_{\bol{i}}$.

\subsection{Description using reactions}
Due to its on-lattice definition, the dynamics of an AIM can be described as a set of reactions between two particle species $A_{\bol{i}}$ and $B_{\bol{i}}$, representing respectively particles at site $\bol{i}$ having $+1$ and $-1$ spin. The model is completely defined once the following two processes are specified:
\begin{enumerate}
    \item[i.] how particles move in space, namely with what rates they undergo hopping reactions
    \begin{equation}
    	A_{\bol{i}}\longrightarrow A_{\bol{j}} \qquad B_{\bol{i}}\longrightarrow B_{\bol{j}}
    \end{equation}
    \item[ii.] how particles change direction, namely with what rate they undergo the spin-flip reactions
    \begin{equation}
    	A_{\bol{i}}\longrightarrow B_{\bol{i}} \qquad B_{\bol{i}}\longrightarrow A_{\bol{i}}
    \end{equation}
\end{enumerate}
We next address these processes in turn.

\subsubsection{Hopping}
In an AIM, particles are assumed to hop with a fixed rate $D$ in all spatial directions, except for the $\hat{x}$ direction where there is a preferred motion set by the spin variable. We thus introduce biased hopping reactions in the $\hat{x}$ direction as
\begin{align}
	A_{\bol{i}}&\longrightarrow A_{\bol{i}\pm \hat{x}}\qquad \text{rate: }D(1\pm \epsilon)
        \label{basatA}\\
        B_{\bol{i}}&\longrightarrow B_{\bol{i}\pm \hat{x}}\qquad \text{rate: }D(1\mp \epsilon)
        \label{Basata}
\end{align}
In all other directions $\hat{y}\neq\hat{x}$ instead, the hopping is unbiased and hence
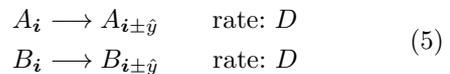
\begin{equation}
\begin{split}
        A_{\bol{i}}&\longrightarrow A_{\bol{i}\pm \hat{y}} \qquad \text{rate: }D\\
        B_{\bol{i}}&\longrightarrow B_{\bol{i}\pm \hat{y}} \qquad \text{rate: }D
\end{split}
\label{diff}
\end{equation}
Here the bias parameter $0\le \epsilon \le 1$ quantifies self-propulsive activity. The hopping reactions are not influenced by the presence of other particles, and hence are independent of particle concentration.

\subsubsection{Spin-flipping}\label{flipper}
In this work we address three different types of AIM (AIM0, AIM1 and AIM2, the latter with several sub-variants), which are distinguished by different choices of spin-flip reaction rates.

{\bf AIM0: Original Ising flip rates}

\noindent In the original formulation of the AIM, as introduced in \cite{solon2013revisiting}, the rates for a spin-flipping event took inspiration from equilibrium dynamics of a fully-connected Ising model in the canonical ensemble. This means that, in absence of any hopping, each {\em site} behaves as a fully-connected Ising model. In terms of reactions between $A$ and $B$ particles, this choice of rates leads to
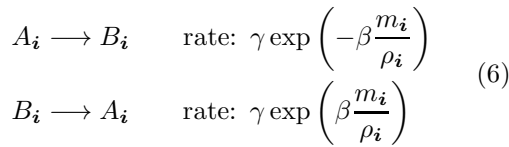
\begin{equation}
\begin{split}
        A_{\bol{i}}&\longrightarrow B_{\bol{i}} \qquad
        \text{rate: }\,\gamma \exp{\left(-\beta\frac{m_{\bol{i}}}{\rho_{\bol{i}}}\right)}
        \\
        B_{\bol{i}}&\longrightarrow A_{\bol{i}} \qquad
        \text{rate: }\,\gamma \exp{\left(\beta\frac{m_{\bol{i}}}{\rho_{\bol{i}}}\right)} 
\end{split}
\label{oAIM}
\end{equation}
which we shall refer to as AIM0.
Here $\gamma$ is the rate of particle flipping in the $m=0$ case, while $\beta$ plays the role of an inverse temperature. 

This choice of flip rates is however unfeasible to implement in a Doi-Peliti framework: although we were able to formally derive a field-theoretical action for this choice, we could not express it in terms of simple functions but only as an infinite series. Given that the choice of rates in \cite{solon2013revisiting} is somewhat arbitrary, we are at liberty to make others for which the field theory is simpler.

{\bf AIM1: Alternative Ising-like flip rates}

\noindent From a technical point of view, what makes it difficult to study the rates of \eqref{oAIM} is the presence of the $\rho_{i}$ in the denominator of the exponential argument. Hence, a choice of reactions which still mimics equilibrium dynamics of Ising spins is given by \cite{kourbane2018exact}
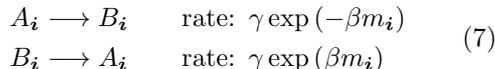
\begin{equation}
\begin{split}
        A_{\bol{i}}&\longrightarrow B_{\bol{i}} \qquad
        \text{rate: }\,\gamma \exp{\left(-\beta m_{\bol{i}}\right)}
        \\
        B_{\bol{i}}&\longrightarrow A_{\bol{i}} \qquad
        \text{rate: }\,\gamma \exp{\left(\beta m_{\bol{i}}\right)} 
\end{split}
\label{aAIM}
\end{equation}
We shall refer to this model as AIM1. The two set of reactions \eqref{oAIM} and \eqref{aAIM} are expected to give qualitatively similar phase diagrams, but quantitative agreement is not expected. In particular, strong differences are expected to emerge in the zero and infinite density limits, where the absence of a normalization of $m_{i}$ by $\rho_{i}$ might lead to drastic consequences. However, we will later show how, at finite densities, the behaviour near the ordering transition is extremely similar.

{\bf AIM2: Collisional flip rates}

\noindent In the context of off-equilibrium systems such as active matter, we have no particular reason to argue that the flip dynamics should mimic that of any equilibrium spin system. The rates that will be introduced here are inspired by the process of multiple-particle collisions, involving a finite and fixed number of particles (chosen at random from the same site), in contrast with the equilibrium-inspired rates, where all particles on the same site interact to set the rates.  We consider the following three reaction processes:

\noindent AIM2.1: One-body collisional flip rate
\begin{equation}
\begin{split}
        A_{\bol{i}}&\longrightarrow B_{\bol{i}} \qquad
        \text{rate: }\,\gamma
        \\
        B_{\bol{i}}&\longrightarrow A_{\bol{i}} \qquad
        \text{rate: }\,\gamma 
\end{split}
\label{gamma}
\end{equation}

\noindent AIM2.2: Two-body collisional flip rate
\begin{equation}
\begin{split}        
A_{\bol{i}}+B_{\bol{i}}&\longrightarrow 2\,B_{\bol{i}} \qquad
        \text{rate: }\,\lambda
        \\
        A_{\bol{i}}+B_{\bol{i}}&\longrightarrow 2\,A_{\bol{i}} \qquad
        \text{rate: }\,\lambda 
\end{split}
\label{lambda}
\end{equation}

\noindent AIM2.3: Three-body collisional flip rate
\begin{equation}
\begin{split}
        2\,A_{\bol{i}}+B_{\bol{i}}&\longrightarrow 3\,A_{\bol{i}} \qquad
        \text{rate: }\,\tau
        \\
        A_{\bol{i}}+2\,B_{\bol{i}}&\longrightarrow 3\,B_{\bol{i}} \qquad
        \text{rate: }\,\tau 
\end{split}
\label{tau}
\end{equation}

\noindent The one-body collision (or random) spin-flipping \eqref{gamma} introduces a random error in the alignment process, not dissimilar to thermal noise. In fact, AIM2.1 is exactly equivalent to the infinite-temperature limit $\beta\to 0$ of both AIM0 \eqref{oAIM} and AIM1 \eqref{aAIM}. It amounts to a random interconversion of $A$ and $B$ particles, and there is no phase transition.

On the other hand, the two- \eqref{lambda} and three-body \eqref{tau} collisional terms favour alignment. For both cases, in the absence of any additional random spin-flipping, the two fully ordered states (all $A$ or all $B$ particles) are absorbing states: once the system reaches them, it will remain there forever. We might therefore expect AIM2.2 and AIM2.3 to give rise to a phenomenology similar to the original AIM0, at least qualitatively, with spontaneous symmetry breaking leading to a strongly flocked state of positive or negative spins. However, in Sec.~\ref{twobody}, we will show how this expectation fails for AIM2.2: the two-body flip reaction cannot create ordering in the presence of any random (one-body) spin-flipping rate, no matter how small. Therefore a three-body interaction (AIM2.3) will be needed below to get an ordering transition. With this term present, one can add back two- and one- body collisional flips without qualitatively altering the outcome; we use the inclusive nomenclature `AIM2' for this most general case.

In the current work we restrict attention to AIM1 and AIM2 as described above. For these (like AIM0) the hopping and spin-flip rules do not obey detailed balance even in this propulsion-free limit ($\epsilon\to 0$) \cite{solon2013revisiting,solon2015flocking}. A different AIM variant was recently constructed specifically to restore detailed balance in this limit \cite{tal2023}, but we do not address it here.

\subsection{Master Equation}\label{MastE}
Having specified the hopping and flip rates, the behaviour of the model can be studied via a Master Equation $\partial_{t}P=\mathcal{L} \left[P\right]$ for the probability distribution $P(\bol{n}^{+},\bol{n}^{-};t)$ in configuration space.  The Master Equation is linear in $P$, and each different process gives an independent contribution to $\mathcal{L}$:
\begin{equation}
	\partial_{t}P=\mathcal{L}_{D}\left[P\right]+\mathcal{L}_{\epsilon}\left[P\right]+\mathcal{L}_{\mathrm{flip}}\left[P\right]\label{eq:MastE}
\end{equation}
where $\mathcal{L}_{\mathrm{flip}}$ is the contribution of the alignment process, $\mathcal{L}_{D}$ arises from the unbiased hopping dynamics while $\mathcal{L}_{\epsilon}$ takes into account the hopping bias and is linear in  the bias parameter $\epsilon$. In the cases of AIM2, the alignment contribution can be further written as $\mathcal{L}_{\mathrm{flip}}=\mathcal{L}_{\gamma}+\mathcal{L}_{\lambda}+\mathcal{L}_{\tau}$, with terms stemming from reaction \eqref{gamma}, \eqref{lambda} and \eqref{tau} respectively. The explicit form of all the evolution operators is given in Appendix \ref{ME}.

\section{The Doi-Peliti field theory}\label{ft}
The Master Equation is {\em exactly} represented by a field-theoretic action \cite{tauber2014critical,wiese2016coherent}, constructed through a coherent-state path integral representation of the evolution operator $\mathcal L$, following the second-quantization formalism to reaction-diffusion processes introduced by Doi \cite{doi1976stochastic,doi1976second} and Peliti \cite{peliti1985path}.

\subsection{Building the action}
Briefly, the Doi-Peliti construction is as follows. For each particle species, {\it creation} and {\it annihilation} fields are introduced. The Master Equation is first written in a second-quantisation formalism, such that the {\it state} of the system -- namely the probability generating function -- evolves via an imaginary-time Schroedinger equation with an evolution operator $\hat H$ derived from $\mathcal L$. The action for the creation and annihilation fields is obtained by computing the matrix elements of $\hat H$ in the basis of the eigenvectors of creation and annihilation operators, of which our fields are the associated eigenvalues. Operationally, one first writes $\hat H$ in normal ordered form, and then replaces annihilation and creation operators with their corresponding fields. See Appendix~\ref{DP} for more details.

\subsection{Building the operators}\label{DPoperators}
The main drawback of the Doi-Peliti formalism is that the fields it describes are of difficult physical interpretation. In fact, not only does the evolution operator have to be written in a second-quantised formalism, but so do the observables of the theory. For example, consider a model with a single species of particles on a lattice. The number of particles on a given site $\bol{i}$ can then be expressed as $n_{\bol{i}}=a_{\bol{i}}^{\dagger}a_{\bol{i}}$, where $a^{\dagger}$ and $a$ are creation and annihilation operators. Say we wanted to compute the expectation value of some observable containing products of $n_{\bol{i}}$ at different sites and times. The rule to construct the corresponding field-theoretical operator is very similar to that needed to build the action. First, particle numbers are written in terms of creation and annihilation operators, then such operators have to be normal ordered, and then operators are substituted by fields. A simplifying feature of the Doi-Peliti theory is that any creation operators appearing at the last of the chosen times can then be dropped. The underlying reason is causality: the event of a particle {\it created} after all the measurements should not affect the averages we are computing.

Accordingly, the field-theoretical operator whose average is equal to the expected value of $n_{\bol{i}}$ at time $t$ is constructed as follows: $n_{\bol{i}}(t)=a_{\bol{i}}^{\dagger}(t)a_{\bol{i}}(t)\rightarrow a_{\bol{i}}(t)\rightarrow \phi_{\bol{i}}(t)$ with $\phi$ the annihilation field. (The creation operator at time $t$ can be dropped as stated above.) Therefore, the following relation for the expected value of $n$ holds
\begin{equation}
	\E{n_{\bol{i}}(t)}=\langle \phi_{\bol{i}}(t)\rangle
\end{equation}
where we denote with $\E{\cdot}$ expected values for the microscopic stochastic process, while $\langle\cdot\rangle$ indicates the average over the field-theoretic measure.
This case is simple, but more complicate operators are not always so intuitive. For example, the correlation between $n_{\bol{i}}$ at time $t$ and $n_{\bol{j}}$ at time $t'<t$  is
\begin{equation}
	\E{n_{\bol{i}}(t)\,n_{\bol{j}}(t')}=\langle \phi_{\bol{i}}(t)\phi_{\bol{j}}^{*}(t')\phi_{\bol{j}}(t') \rangle
		\label{corrDP}
\end{equation}
where $\phi^*$ is the creation field. Meanwhile the equal time and equal position correlator obeys
\begin{equation}
	\E{n_{\bol{i}}(t)^{2}}=\langle \phi_{\bol{i}}(t)^{2} + \phi_{\bol{i}}(t)\rangle
	\label{corrDP2}
\end{equation}
This follows from normal ordering whereby 
$$n^{2} \to \left(a^{\dagger}_{\bol{i}} a_{\bol{i}}\right)^{2} = a^{\dagger}_{\bol{i}} a^{\dagger}_{\bol{i}} a_{\bol{i}}\, a_{\bol{i}}+a^{\dagger}_{\bol{i}}a_{\bol{i}} \to\phi^2_{\bol{i}}+\phi_{\bol{i}}\, .$$

\subsection{The Doi-Peliti action for AIMs}
Active Ising Models have two distinct particle types $A,B$ corresponding to spins $\pm 1$ respectively, so alongside annihilation and creation fields $\phi$, $\phi^{*}$ for species $A$ we need counterparts $\psi$ and $\psi^{*}$ for $B$. Just as for the Master Equation, the spacetime action $S$ of the field theory is additive over the various hopping and jump processes, and also over spatial (site) and temporal variables. Thus $S=\sum_{\bol{i}}\int dt \,\mathcal{S}$ with the action density
\begin{multline}
	\mathcal{S} = \phi_{\bol{i}}^*(t)\partial_t \phi_{\bol{i}}(t) + \psi_{\bol{i}}^*(t)\partial_t \psi_{\bol{i}}(t) +\\ +\mathcal{S}_{D}+\mathcal{S}_{\epsilon}+\mathcal{S}_{\mathrm{flip}}
\label{DPaction}
\end{multline}
Since the spin-flip dynamics involves only same-site particles, $\mathcal{S}_{\mathrm{flip}}$ is fully local in both space and time, while the diffusive $\mathcal{S}_{D}$ and propulsive $\mathcal{S}_{\epsilon}$ hopping contributions connect neighbouring sites. The explicit form of these various contributions for the different AIMs is given in Appendix \ref{ciack}.

\subsection{The Cole-Hopf transformation}\label{ch}
The Cole-Hopf transformation \cite{tauber2014critical,lefevre2007dynamics} connects the somewhat abstract Doi-Peliti fields to physical observables, namely number-density fields for $A$ and $B$ particles. For the one-species example of Sec \ref{DPoperators}, the transformed fields $\rho$ and $\tilde \rho$ obey
\begin{equation}
\phi^{*}=e^{\tilde{\rho}}\, , \qquad \phi=e^{-\tilde{\rho}}\rho
\label{CHange}
\end{equation}
Thus the density field $\rho=\phi^{*}\phi$ is analogue to the second-quantised number operator $\hat{n}=a^{\dagger}a$, while the correlation function of Eq.~\eqref{corrDP} now takes the more intuitive form
\begin{equation}
	\E{n_{\bol{i}}(t)\,n_{\bol{j}}(t')}=\langle \rho_{\bol{i}}(t)\rho_{\bol{j}}(t') \rangle
\end{equation}
More generally, for all density correlators evaluated at different times and/or different sites, one can now replace the expectation value by the average over the field-theoretical measure, and replace the particle number operators by the corresponding $\rho$ fields.

However, to compute correlation functions on the same site at the same time, subtleties remain, because the corresponding number operators must remain normal-ordered. Thus the correlator given in Eq.~\eqref{corrDP2} obeys $\E{n_{\bol{i}}(t)^{2}} = \langle \rho_{\bol{i}}(t)^{2} + \rho_{\bol{i}}(t)\rangle$. This non-intuitive result is the unavoidable price for building an exact theory in terms of (almost!) physical density fields. Below we therefore pay careful attention when computing equal-time correlators.

\section{The hydrodynamic limit}\label{hydro}
Here we derive hydrodynamic-level equations for the various Active Ising Models proposed in Sec~\ref{aimdef}. (We exclude AIM0 because, as mentioned there, its Doi-Peliti action is intractable.) The derivation is lengthy, but offers important insights. The strategy is as follows: starting from the Master Equation we derive the Doi-Peliti field theory following Sec~\ref{ft}. Converting to physical fields via Cole-Hopf (as in Sec~\ref{ch}), we use a reverse Martin-Siggia-Rose procedure (see Appendix~\ref{MSR}) to derive, from the field-theory action, equations of motion for the density fields. This programme can be followed exactly to the last stage, at which point the non-Gaussian noise that emerges at exact level (see Appendix~\ref{MSR}) can be either gaussianized (to give the Langevin equations) or suppressed (to give deterministic hydrodynamics). The last stage is achieved by sending the linear size of the system $L\to\infty$ while keeping fixed the density of particles. In this limit, exact hydrodynamic PDEs emerge, describing the behaviour of hydrodynamic variables on scales comparable with $L$, while the leading order stochastic corrections give the Gaussian (Langevin) noises.

\subsection{Preliminaries}
The Cole-Hopf transformed action density reads
\begin{equation}
\mathcal{S}=\tilde{\rho}_{\bol{i}}^{+}\partial_{t}\rho_{\bol{i}}^{+}+\tilde{\rho}_{\bol{i}}^{-}\partial_{t}\rho_{\bol{i}}^{-}+\mathcal{S}_{D}^{CH}+\mathcal{S}_{\epsilon}^{CH}+\mathcal{S}_{\mathrm{flip}}^{CH}
\end{equation}
where $\tilde{\rho}^{+}$ and $\rho^{+}$ have replaced $\phi$ and $\phi^{*}$, and $\tilde{\rho}^{-}$ and $\rho^{-}$ have replaced $\psi$ and $\psi^{*}$. The fields $\rho^{+}$ and $\rho^{-}$ approach the physical densities for $A$ and $B$ particles respectively. The contributions to $\mathcal{S}^{CH}$ are found via the change of variables \eqref{CHange}; their forms are given as needed, below.

We first set ({\em wlog}) the lattice spacing to $h=1$, and then consider the system at diffusive hydrodynamic scales, achieved by a further rescaling of spatial coordinates, $\tilde{\bol{x}}=\bol{i} / L$, and of time, $\tilde{t}=t/L^{2}$. This choice of rescaling follows from requiring diffusion to be the process that fixes the hydrodynamic time-scale. Under these rescalings, we have
$
\sum_{\bol{i}}=L^{d}\int d\tilde{\bol{x}} \,;\, \int d t = L^{2}\int d\tilde t
$, and  $S=\int d\tilde{\bol{x}} d\tilde t\,\tilde{\mathcal{S}}$, where $\tilde{\mathcal{S}}$ is the hydrodynamic action density, which absorbs all the powers of $L$ coming from space-time rescaling. This action density can be expanded in powers of $L^{-1}$, dropping subleading terms as $L\to\infty$. We continue to split $\tilde{\mathcal{S}}$ into contributions from spin-flip, diffusive and biased hopping processes, whose rates must however be rescaled such that all three contribute in the hydrodynamic limit. Finally, the conjugate fields must also be rescaled as $\tilde \rho\to L^{-d}\tilde \rho$.

\subsection{Hydrodynamics for AIM1}
For AIM1, with spin-flip rates given by \eqref{aAIM}, the deterministic hydrodynamic equations are known from Ref.~\cite{kourbane2018exact}, offering an important cross check on our methods. At leading order in $L^{-1}$, the action terms (dropping the $CH$ superscript) are:

\begin{equation}
\begin{split}
	\tilde{\mathcal{S}}_{D}&=-L^{d} D \tilde\rho^{+} \tilde{\nabla}^{2} \rho^{+} -L^{d} D \tilde\rho^{-} \tilde{\nabla}^{2} \rho^{-}-\\
	&- L^{d} D \rho^{+}\left(\tilde{\bol{\nabla}} \tilde{\rho}^{+}\right)^{2} - L^{d} D \rho^{-}\left(\tilde{\bol{\nabla}} \tilde{\rho}^{-}\right)^{2}
\end{split}
\label{diff1}
\end{equation}
\begin{equation}
	\tilde{\mathcal{S}}_{\epsilon}=L^{d+1} v \tilde\rho^{+}\partial_{\hat{\tilde{x}}}\rho^{+} - L^{d+1} v \tilde\rho^{-}\partial_{\hat{\tilde{x}}}\rho^{-}
\label{active1}
\end{equation}
\begin{equation}
\begin{split}
	\tilde{\mathcal{S}}_{\mathrm{flip}}&=L^{d+2}\gamma\, e^{-\beta }  \left(e^{\tilde{\rho}^{+}}-e^{\tilde{\rho}^{-}}\right) \times\\
	\times&\left(e^{-\tilde{\rho}^{+}}\rho^{+}  e^{\left(e^{\beta }-1\right) \rho^{-} +\left(e^{-\beta }-1\right) \rho^{+} }-\right.\\
	-&\left.e^{-\tilde{\rho}^{-}}\rho^{-}  e^{\left(e^{-\beta }-1\right) \rho^{-} +\left(e^{\beta }-1\right) \rho^{+} }\right)
\end{split}
\label{flip1}
\end{equation}

For all three to contribute in the hydrodynamic limit, as previously discussed, then if $D$ is fixed of order unity, we must choose $\gamma\sim L^{-2}$  in \eqref{flip1} and $v\sim L^{-1}$  in \eqref{active1}, and redefine these parameters now to absorb such factors. These choices ensure that the number of spin flips is order one in the time $\sim L^2/D$ needed for a particle to diffuse a distance $L$, and that propulsion likewise competes with both flipping and diffusion at this hydrodynamic scale. After the final rescaling mentioned above, $\tilde \rho\to L^{-d}\tilde \rho$, we can look at all terms in $\tilde {\mathcal{S}}$ (including the time derivative terms) scaling as $L^0$, as is required for the $L\to\infty$ limit to now be taken. We finally get to the hydrodynamic action density

\begin{equation}
\begin{split}
	\tilde{\mathcal{S}}&=\tilde\rho^{+}\left(\partial_{\tilde{t}} -D \tilde{\nabla}^{2} + v \partial_{\hat{\tilde{x}}}\right)\rho^{+} +\\
	&+ \tilde\rho^{-}\left(\partial_{\tilde{t}} -D \tilde{\nabla}^{2} - v \partial_{\hat{\tilde{x}}}\right)\rho^{-}  +\\
	&+\gamma\, e^{-\beta }  \left(\tilde{\rho}^{+}-\tilde{\rho}^{-}\right) \times\\
	&\times\left(\rho^{+}  e^{\left(e^{\beta }-1\right) \rho^{-} +\left(e^{-\beta }-1\right) \rho^{+} }-\right.\\
	&\quad-\left.\rho^{-}  e^{\left(e^{-\beta }-1\right) \rho^{-} +\left(e^{\beta }-1\right) \rho^{+} }\right)
\end{split}
\label{mannaggia1}
\end{equation}

The absence of higher powers of the $\tilde\rho$ fields finally allows us to map this field theory, via the inverse Martin-Sigga-Rose procedure outlined in Appendix~\ref{MSR}, onto the noiseless limit of a set of stochastic PDEs (the noisy version is given in Sec \ref{flu} below). The hydrodynamic equations governing $\rho^{+}$ and $\rho^{-}$ are thereby found as
\begin{align}
	\partial_{t} \rho^{+}&=D\nabla^{2}\rho^{+}- v\partial_{\hat x}\rho^{+}- F(\rho^{+},\rho^{-})\label{paAIM}\\
	\partial_{t} \rho^{-}&=D\nabla^{2}\rho^{-}+v\partial_{\hat x}\rho^{-} + F(\rho^{+},\rho^{-})\label{maAIM}
\end{align}
where  \begin{equation}
\begin{split}
	F(\rho^{+},\rho^{-})=\gamma\, e^{-\beta }&\left(\rho^{+}  e^{\left(e^{\beta }-1\right) \rho^{-} +\left(e^{-\beta }-1\right) \rho^{+} }-\right.\\
	&-\left. \rho^{-}  e^{\left(e^{-\beta }-1\right) \rho^{-} +\left(e^{\beta }-1\right) \rho^{+} }\right)\nonumber
\end{split}
\end{equation}
If written in terms of magnetisation $m=\rho^{+}-\rho^{-}$ and total number of particles $\rho=\rho^{+}+\rho^{-}$, these equations become
\begin{align}
	\partial_{t} m&=D\nabla^{2}m-v\partial_{\hat x}\rho- 2 F(m,\rho)\label{mimuovo}\\
	\partial_{t} \rho&=D\nabla^{2}\rho-v\partial_{\hat x}m\label{rimuovo}
\end{align}
where 
\begin{equation}
\begin{split}
    F&\left(m,\rho\right)=\gamma\, e^{-\beta -\rho+\rho \cosh\beta}\times\\
        &\times( m \cosh{\left[ m \sinh\beta\right] }-\rho \sinh{\left[m \sinh\beta\right]})
\end{split}
\label{fortissima}
\end{equation}
Notably, the `aligning force' $F$ is exactly as found in Ref.~\cite{kourbane2018exact}. There the hydrodynamic equations were derived directly by averaging the microscopic process over a local Possion measure. Although the derivation is quite different, the Doi-Peliti formalism ultimately gives an equivalent result because it is constructed from coherent states that also correspond to a Poisson distribution \cite{tauber2014critical}.

\subsubsection{Fluctuating hydrodynamics}\label{flu}
An advantage of our Doi-Peliti field theory is that it provides a systematic way to address  {\em fluctuating} hydrodynamics. This can be done by keeping the next order in $L^{-d}$ beyond the action \eqref{mannaggia1}. This captures for finite size systems the leading order (small, Gaussian) fluctuations around Equations \eqref{mimuovo}, \eqref{rimuovo}, by adding to them Langevin noises scaling as $L^{-d/2}$. Adding these terms to the action \eqref{mannaggia1}, the equations for $m$ and $\rho$ become
\begin{align}
	\partial_{t} m&=D\nabla^{2}m-v\partial_{\hat x}\rho- 2 F(m,\rho)+ \frac{1}{\sqrt{L^d}}\theta\, \label{mifluttuo}\\
	\partial_{t} \rho&=D\nabla^{2}\rho-v\partial_{\hat x}m+ \frac{1}{\sqrt{L^d}}\bol{\nabla}\cdot\bol{\zeta}\, \label{rifluttuo}
\end{align}
where $F(m,\rho)$ is still given by \eqref{fortissima}, but now we have the noise contributions $\theta$ and $\bol{\zeta}$. The noise $\theta$ can be further split in two contributions $\theta=\eta+\bol{\nabla}\cdot\bol{\xi}$, where the latter arises from diffusion and thus conserves the total magnetisation. The statistics of these Gaussian noises is fully determined by a covariance matrix comprising
\begin{equation}
    \begin{split}
    \langle\eta(\bol{x},t)\eta(\bol{y},s)\rangle&= \,4\,\gamma\,
    e^{-\beta -\rho+\rho \cosh(\beta)}\,\times\\
    &\times\left(
    \rho \cosh{\left[ m \sinh(\beta)\right]}-\right.
    \\
    &-\left.
    m \sinh{\left[m \sinh(\beta)\right]}
    \right)\times\\
    &\times
    \delta\left(\bol{x}-\bol{y}\right)\delta\left(t-s\right)\, \nonumber
    \end{split}
\end{equation}
with other noise covariances being zero except for
\begin{align}
    \langle\xi_i(\bol{x},t)\xi_j(\bol{y},s)\rangle &=
    2\,D\,\rho\,
    \delta_{i,j}
    \delta\left(\bol{x}-\bol{y}\right)
    \delta\left(t-s\right)\, \nonumber \\
    \langle\zeta_i(\bol{x},t)\zeta_j(\bol{y},s)\rangle &=
    2\,D\,\rho\,
    \delta_{i,j}
    \delta\left(\bol{x}-\bol{y}\right)
    \delta\left(t-s\right)\, \nonumber \\
    \langle\xi_i(\bol{x},t)\zeta_j(\bol{y},s)\rangle &=
    2\,D\,m\,
    \delta_{i,j}
    \delta\left(\bol{x}-\bol{y}\right)
    \delta\left(t-s\right)\, \nonumber
\end{align}
Note that $\bol{\xi}$ and $\bol{\zeta}$, namely the conservative noises, are gaussian also beyond the large $L$ limit. This can be seen from the fact that they arise from the action terms \eqref{diff1} and \eqref{active1}, where no term is more than quadratic in $\tilde{\rho}^{\pm}$. The non-conservative noise $\eta$, on the other hand, has a non-gaussian statistics (higher powers of $\tilde{\rho}^{\pm}$ in \eqref{flip1}) which becomes gaussian only al large $L$ in virtue of the central limit theorem.

\subsection{Hydrodynamics for AIM2}\label{HAIM2}
The same procedure as used above for flip rates obeying \eqref{aAIM} can be applied to the many-body rates \eqref{gamma}-\eqref{tau}. Spatial hopping is not affected, so all the contributions proportional to $D$ and $\epsilon$ will remain unchanged.  But the contribution $\tilde{\mathcal{S}}_{\mathrm{flip}}$ to the hydrodynamic action now takes the form (before rescaling parameters)
\begin{equation}
\begin{split}
	\tilde{\mathcal{S}}_{\mathrm{flip}}=&L^{d+2}\gamma\, \left(e^{\tilde{\rho}^{+}}-e^{\tilde{\rho}^{-}}\right) \times\\
	&\times\left(e^{-\tilde{\rho}^{+}}\rho^{+}  - e^{-\tilde{\rho}^{-}}\rho^{-} \right)-\\
	-& L^{d+2}\,\lambda\, \left(e^{\tilde \rho^{+}}-e^{\tilde \rho^{-}}\right)^2  \times\\
	&\times e^{-\tilde \rho^{+}-\tilde \rho^{-}}  \rho^{+}\, \rho^{-} +\\
	+&L^{d+2} \frac{\tau}{2}  \left(e^{-\tilde \rho^{+}}-e^{\tilde \rho^{-}}\right) \times\\
	&\times \left(e^{\tilde \rho^{+}} \rho^{+} -  e^{\tilde \rho^{-}} \rho^{+}\right) \rho^{+}\, \rho^{-}\, 
\end{split}
\label{SaAIM}
\end{equation}
As done previously, we now rescale the rates $\gamma$, $\lambda$ and $\tau$ by $L^{-2}$ such that each type of flip occurs competes with diffusion (and propulsion). Finally rescaling again $\tilde \rho\to L^{-d}\tilde \rho$ and taking $L\to\infty$, the resulting hydrodynamic action becomes equivalent to the same partial differential equations \eqref{maAIM}, \eqref{paAIM}, but with a different choice of $F(\rho^{+},\rho^{-})$. Again rewriting this in terms of magnetisation $m=\rho^{+}-\rho^{-}$ and particle density $\rho=\rho^{+}+\rho^{-}$, we recover \eqref{mimuovo} and \eqref{rimuovo}, with \eqref{fortissima} replaced by 
\begin{equation}
    F\left(m,\rho\right)=m \left( \gamma +\tau \frac{m^2-\rho ^2}{8}  \right)\, 
\label{fortissima2}
\end{equation}
Just as in Sec~\ref{flu}, we can compute leading-order fluctuation corrections, recovering \eqref{mifluttuo} and \eqref{rifluttuo}, in which  $F(m,\rho)$ obeys \eqref{fortissima2} and the noise correlator of $\eta$ given by
\begin{equation}
    \begin{split}
    \langle\eta(\bol{x},t)&\eta(\bol{y},s)\rangle= \delta\left(\bol{x}-\bol{y}\right)\delta\left(t-s\right) \times\\
    &\times\Bigl[2\gamma\rho+\left(\lambda +\rho  \frac{\tau}{4} \right)(\rho^{2} -m^{2})\Bigr]
    \end{split}\label{NAIM2}
\end{equation}
while all other correlators remain the same.

\subsection{Homogeneous solutions}
Spatially homogeneous but time-dependent solutions of the noiseless  hydrodynamic equations are found by assuming $m\left(\bol{x},t\right)= m(t)$ and $\rho\left(\bol{x},t\right)= \rho(t)$ in (\ref{mimuovo},\ref{rimuovo}), which become
\begin{align}
	\partial_{t} m&=- 2 F(m,\rho)\, \label{linearem}\\
	\partial_{t} \rho&=0\, 
\end{align}
The second of these expresses particle conservation: $\rho(t) = \rho_0$, the initial density. In contrast, $m$ relaxes via the spin-flip dynamics, with an asymptotic solution $\lim_{t\to0} m(t)=m_{0}$ obeying $F(m_{0},\rho_{0})=0$. For both choices of $F$ considered above in \eqref{fortissima}, \eqref{fortissima2}, $m_{0}=0$ is always a solution but is unstable if $\partial_{m}F(m_{0},\rho_{0})<0$, giving a magnetized phase. For definiteness we focus on AIM2 here (though AIM1 is similar~\cite{kourbane2018exact})
for which the force $F(m,\rho)$  obeys \eqref{fortissima2} so that
\begin{equation}
\partial_{m} F=\left(\gamma-\frac{\tau}{8}\rho^{2}\right)-\frac{3 \tau}{8}m^{2}
\end{equation}
The state $m_{0}=0$ is thus stable for  $\rho_{0}\leq\rho_{c}=(8\gamma/\tau)^{1/2}$, and unstable for $\rho_{0}>\rho_{c}$, where one has a symmetric pair of stable, magnetized states $m_0=\pm \bar m$ with $\bar m^2=\rho_{0}^{2}-\rho_{c}^{2}$. This resembles a standard, Ising-like spontaneous symmetry breaking where two vanishingly magnetic states merge at the critical point $\rho_{0}=\rho_{c}$. However, in the passive Ising model, for all $\rho_{0}>\rho_{c}$ the two solutions $m = \pm m_{0}=\pm \bar m$ remain stable against {\em inhomogeneous perturbations}. For AIMs this is not the case: there is a region of parameter space where no homogeneous solution is stable. The AIM transition is thus better understood as a first-order transition, akin to a liquid-gas transition \cite{solon2015flocking}.

\subsubsection{Linear stability of uniform states}\label{lsa}
To check the linear stability of homogeneous solutions $m=m_0$, $\rho=\rho_0$, we linearise the equations of motion and examine small perturbations $\delta m$ and $\delta \rho$ which then obey:
\begin{equation}
\begin{split}
	\partial_t \delta m&=D\,\nabla^2 \delta m - v\,\partial_x \delta \rho\\
	 &-2\,\alpha\left(\rho_0\right) \delta m -2\,g\left(\rho_0\right)\delta \rho
\end{split} 
\label{hydrom00}
\end{equation}
\begin{equation}
	\partial_t\delta \rho=D\,\nabla^2\delta\rho - v\,\partial_x\delta m \label{hydrorho00}
\end{equation}
where
\begin{align}
    \alpha\left(\rho_0\right)&= \partial_m F\left(m_{0},\rho_{0}\right)\\
    g\left(\rho_0\right)&= \partial_\rho F\left(m_{0},\rho_{0}\right)
\end{align}
In Fourier space 
($ f\left(\bol{k},t\right)=\int d^dx\, f\left(\bol{x},t\right) e^{-\iu\bol{x}\cdot\bol{k}}$), the linearised dynamics becomes
\begin{equation}
    \partial_t
    \left(
    \begin{array}{cc}
    \delta m \\
    \delta \rho\\
    \end{array}
    \right)=
    M\left(\bol{k}\right)
    \left(
    \begin{array}{cc}
    \delta m \\
    \delta\rho\\
    \end{array}
    \right)
\end{equation}
Here
\begin{equation}
    M(\bol{k})=
    \left(
    \begin{array}{cc}
    - i\, v\, k_x -2\,g(\rho_0)\;&\; -D\, k^2-2\,\alpha(\rho_0) \\
    -D\, k^2 & - i\, v\, k_x 
    \end{array}
    \right)
\end{equation}
and stability against perturbations at wavevector $\bol{k}$ requires both eigenvalues of $M(\bol{k})$ to have a nonpositive real part. 
These eigenvalues  are
\begin{equation}
\begin{split}
    \lambda_{1} \left(\bol{k}\right)=&-\sqrt{\alpha(\rho_0)^2 +2 i \, v\,  g(\rho_0) k_x-v^2 k_x^2 }\\
    &- \alpha(\rho_0) -D k^2
\end{split}
\end{equation}
\begin{equation}
\begin{split}
    \lambda_{2} \left(\bol{k}\right)=&\sqrt{\alpha(\rho_0)^2 +2 i \, v\,  g(\rho_0) k_x-v^2 k_x^2 }\\
    &- \alpha(\rho_0) -D k^2
\end{split}
\end{equation}
Studying the eigenvalues at $k=0$ (where $\lambda_1 = -2 \alpha(\rho_0)$ and $\lambda_2 = 0$) we confirm the analysis made above concerning stability within the subspace of homogeneous (mean-field) solutions.

What happens if instead we perturb the system, not with a homogeneous perturbation, but with a slowly varying one? For a system with finite positive $\alpha(\rho_0)$ (hence stable against uniform perturbations) continuity in $\bol{k}$ requires $\Re\left(\lambda_1(\bol{k})\right)<0$ at small $\bol{k}$. In contrast,  $\lambda_2$ at small $\bol{k}$ takes the form
\begin{equation}
\begin{split}
	\lambda_2 \left(\bol{k}\right)=& - D k^2 + v^2\frac{ g\left(\rho_0\right)^2-\alpha\left(\rho_0\right) ^2}{2 \, \alpha\left(\rho_0\right)^3} k_x^2 +\\
	& + i \frac{ v\, g\left(\rho_0\right)}{\alpha\left(\rho_0\right)} k_x  + O(k^3)
\end{split}
\end{equation}
We distinguish the cases $\rho_{0}<\rho_{c}$ and $\rho_{0}>\rho_{c}$:
\begin{enumerate}
    \item At $\rho_{0}<\rho_{c}$, the only solution is $m_0=0$, $g\left(\rho_0\right)=0$ and $\alpha\left(\rho_0\right)=\frac{\tau}{8}\left(\rho_{c}^{2}-\rho_{0}^{2}\right)>0$. Hence, the eigenvalue $\lambda_2$ becomes
\begin{equation}
    \lambda_2 \left(\bol{k}\right)= -D k^2-\frac{v^2}{2\alpha(\rho_{0})} k_x^2 + O(k^3)
\end{equation}
indicating stability of the uniform, nonmagnetic solution for all $\rho_{0}<\rho_{c}$, in agreement with the predictions of mean field theory.
\item At $\rho_{0}>\rho_{c}$ the homogeneous solutions that appear stable from a mean-field argument have $m_0^{2}=\rho_{0}^{2}-\rho_{c}^{2}\neq 0$. 
In this case, we have that $\alpha(\rho_{0})=\frac{\tau}{4} \left(\rho_{0}^{2}-\rho_{c}^{2}\right)>0$ 
and $g(\rho_{0})= \mp \frac{\tau \rho_{0}}{4} \left(\rho_{0}^2 - \rho_{c}^2\right)^{1/2}$. 
Therefore,
\begin{equation}
\begin{split}
	\lambda_2 \left(\bol{k}\right)&= - D k^2 + v^2\frac{ 2 \rho_{c}^{2}}{\tau\left(\rho_{0}^{2}-\rho_{c}^{2}\right)^{2}} k_x^2 +\\
	&\mp i \frac{ \rho_{0} \,v}{\left(\rho_{0}^2 - \rho_{c}^2\right)^{1/2}} k_x + O(k^3)
\end{split}
\end{equation}
The linear part (in $\bol{k}$) of $\lambda_2 \left(\bol{k}\right)$ is always imaginary, and hence does not affect the stability analysis. The quadratic part may, however, become positive for values of $\rho_0$ close to $\rho_{c}$. In particular, this happens when
\begin{equation}
	\rho_{0}^{2}-\rho_{c}^{2} < \sqrt{\frac{2}{\tau D}}\,\rho_{c}\, v \Rightarrow
\end{equation}
\begin{equation}
\begin{split}
	\rho_{0}<\rho_{l}&:=\sqrt{\rho_{c}^{2}+ \sqrt{\frac{2}{\tau D}}\,\rho_{c}\, v}\;=\\
	&=\rho_{c}+\frac{v}{\sqrt{2\tau D}}+\mathcal{O}(v^{2})
\end{split}
\end{equation}
\end{enumerate}
In this second scenario, which arises for nonzero propulsion $v$, the homogeneous magnetic phase becomes unstable with respect to long wavelength perturbations.  Only for $v = 0$ is the passive-Ising-like second order transition recovered; for all $v\neq0$ there is a range of densities, $\rho_{c}(\gamma,\tau)<\rho_{0}<\rho_{l}(\gamma,\tau,D)$, in which no homogeneous solution is stable. In this range the system is therefore driven towards a spatiotemporal pattern. 

Although we will not reproduce here the full calculation, note that the same qualitative behaviour arises for AIM1, in which the force $F$ in \eqref{linearem} is replaced by \eqref{fortissima}: here it is again possible to show that for $v\neq0$ there is a finite range of densities $\rho_{c}<\rho_{0}<\rho_{l}$ in which the ordered homogeneous solution is linearly unstable with respect to long-wavelength spatial perturbations. Hence, the transition is not second-order, but is better understood as a liquid-gas phase transition as in \cite{solon2015flocking}. In both cases, for the zero propulsion limit $v\to0$, we find $\rho_{l}\to\rho_{c}$, so that the homogeneous ordered and disordered phases are linearly stable on either side of $\rho_c$, and we predict a second-order transition in that limit.

\section{Role of two-body collisions}\label{twobody}
In the previous section we analysed the hydrodynamic behaviour of AIM2, where the spin flipping process was given by the set of reactions (\ref{lambda},\ref{tau}). Strikingly, the critical density $\rho_c = (8\gamma/\tau)^{1/2}$ depends on the one-body (random) spin-flip rate $\gamma$, and the three-body rate $\tau$, but not on the two-body rate $\lambda$. This means that, contrary to naive expectation, two-body collisional alignment cannot by itself lead to ordering, no matter how large the rate $\lambda$ at which this occurs.

A physical interpretation of the relevant process is that two close enough particles, {\em i.e.} sharing the same lattice site, bump into each other with some rate $\lambda$. When such a collision occurs, if the particles have opposite spin, they align (randomly choosing which of the two orientations to share). Since the spin sets the preferred direction of motion of the particle, the two colliding particles move in the same direction after the collision. This seems to capture a basic and intuitive mechanism through which flocking might occur, yet we find no ordered phase. Something closer to a `majority rule' (which gets encoded in the three-body collision rate $\tau$) is instead required.

Intriguingly, several recent studies have proposed that two-body interactions are indeed not enough to sustain global alignment \cite{chatterjee2019threebody,suzuki2015polar,jhawar2020noise}. Our work confirms this prediction, which we believe has not been given enough emphasis in the community. The advantage of our field-theoretical approach is that our exact analysis can cleanly and unambiguously rule out any ordered state induced by the two-body collision term in the hydrodynamic limit addressed here. Specifically, if we retain only the one-body (randomizing) and two-body terms in by setting $\tau=0$ in AIM2, we obtain (\ref{mimuovo},\ref{rimuovo}) with a force term $F(m,\rho) = \gamma m$. The homogeneous solution at zero magnetization, $m_0 =0$, is then stable for all $\gamma>0$, regardless of the global density $\rho_0$. Therefore, for any finite amount of random spin flipping, the two-body collision process described by the reaction \eqref{lambda} is not sufficient to induce collective motion.

At $\gamma=0$, things look slightly different.  Without the two-body term ($\lambda = 0$), all solutions can be written as a superposition of waves which travel in the $\pm x$ direction with speed $v$ and damping $D k^{2}$. These solutions not only conserve the total density, but also the total magnetisation; accordingly a state of uniform magnetization cannot emerge from an unmagnetized initial state. Remarkably, this result is sustained, at hydrodynamic level, even when the two-particle interaction \eqref{lambda} is switched on. This result seems counter-intuitive. Indeed, in the absence of random spin-flipping but with two-body collisions ($\gamma = 0, \lambda >0$) the system has two absorbing states: whenever particles are all either of the $A$ or $B$ kind, no further spin flipping can occur. Either state would represent a permanently stable flock. 

As we have seen, this physics does not emerge in the hydrodynamic limit; we now ask why. A key factor will be that absorbing states are reached in a finite time only in finite-size system. We must therefore switch attention to the fluctuating hydrodynamics of this system arising at finite $L$.

The finite-size behaviour of the two-particle interaction model, at large $L$, is given by
\begin{align}
	\partial_{t} m&=D\nabla^{2}m-v\partial_{\hat x}\rho+ \frac{1}{\sqrt{L^d}}\left(\eta+\bol{\nabla}\cdot\bol{\xi}\right)\, \\
	\partial_{t} \rho&=D\nabla^{2}\rho-v\partial_{\hat x}m+ \frac{1}{\sqrt{L^d}}\bol{\nabla}\cdot\bol{\zeta}\, 
\end{align}
We have already set $\gamma=0$, so this is `pure' AIM2.2 as defined by \eqref{lambda}. As in the previous models, $\eta$, $\bol{\xi}$ and $\bol{\zeta}$ are Gaussian noises whose correlators are found by setting $\gamma = \tau = 0$ in the more general results given already for AIM2 in Sec \ref{HAIM2}.

The noises $\bol{\xi}$ and $\bol{\zeta}$ arise from the diffusive motion of particles, and hence conserve the total magnetisation. Flocking, were it to emerge, would have to stem from the $\eta$ noise term. But, as seen from the covariance results in Sec \ref{HAIM2}, specifically \eqref{NAIM2}, the noise $\eta$ is larger the smaller the magnetisation. When $m\sim 0$, this noise therefore pushes the system towards magnetised states with $m\neq0$. The noise then weakens, so it is less likely for the system to return to $m\sim 0$. When eventually the system reaches the absorbing state $m=\pm\rho$, all particles flock in the same direction forever after. The $\eta$ term therefore does push the system towards a flocking state; but it is the only term that does so. This means that for AIM2.2 any collective motion arises by a purely stochastic mechanism, not a deterministic drift -- a fact also clear from the shape of $F(m,\rho)$ when $\tau = 0$.
As previously discussed, stochasticity, and hence the probability of achieving this flocked state, vanishes in the hydrodynamic limit $L\to\infty$. Therefore, exact conservation of the total magnetisation at deterministic level in is not because spin-flipping processes are absent altogether, but because the probability of having a fluctuation that macroscopically changes $m$ vanishes when $L\to\infty$.  This peculiar scenario is of course radically changed by the three-body collisional coupling term $\tau$, which restores a deterministic drift towards flocking that wins out above the critical density $\rho_c$. 

\section{The AIM critical point}\label{trans}
The linear stability analysis performed in Sec \ref{lsa} shows that AIMs generically undergo a first order transition, with a continuous transition recovered in the limit of unbiased hopping rates $\epsilon\to0$ (equivalently $v\to 0$): this accordingly defines the AIM critical point.
An important question concerns the universality class of this critical transition. The answer would be obvious if this limit recovered a reversible model, which would surely lie in the kinetic Ising class known as Model C \cite{hohenberg1977theory}, as discussed further below. Indeed,
 numerical simulations in $2$ dimensions of the AIM0 give results compatible with this prediction \cite{solon2015flocking}. However, this outcome is not guaranteed because, as also shown in \cite{solon2015flocking}, the dynamics of  AIM0 violates detailed balance even at $v=0$, making the system out of equilibrium even in the absence of self-propulsion. This is equally true of AIM1 and AIM2, and given their shared symmetries one can expect all these models to lie in a single universality class (that may or may not be that of equilibrum Model C). 

A major advantage of our field-theoretic approach is that it creates a clear and unambiguous foundation for resolving this issue via a full renormalization group (RG) analysis. Such an analysis lies beyond our present scope and will be presented elsewhere \cite{futurepaper}. Here we derive a suitable starting point for RG calculations, compare it with the corresponding Model C equations, and review what is known about the two cases. Our starting point is AIM2 where spin-flipping is given by the reactions \eqref{gamma} and \eqref{tau}. We set the two-body collision term \eqref{lambda} to zero but have checked that the results below are unchanged by this, and also checked that they hold for AIM1 with rates \eqref{aAIM}. 

\subsection{Relevant and irrelevant terms}
The hydrodynamic methods used in Sec \ref{hydro} generally identify a limit in which noiseless, mean-field critical behaviour is recovered; this approach does not capture all relevant terms for RG purposes. To identify these, we start instead from a coarse-grained continuous version of the microscopic theory, describing the system on mesoscopic scales (much larger than $h$, the lattice spacing, and much smaller than $L$, the system size). We are hence not assuming anymore the scaling with $L$ of the coefficients investigated in Sec.~\ref{hydro}. We will instead take a continuum limit by sending the lattice spacing $h\to0$. The continuum limit therefore represents a way to investigate the dynamics on scales much larger than $h$, but yet much smaller than $L$. To take the continuum limit $h\to0$, we must then appropriately rescale the hopping and flipping rates and also the particle density fields; see  Appendix~\ref{cont}.

After these rescalings, the action becomes 
$
	S=\int \left(\mathcal{S}_{D}+\mathcal{S}_{\mathrm{flip}}\right) \,d\bol{x}dt
$
where
\begin{equation}
\begin{split}
	\mathcal{S}_{D}&=- D \tilde\rho^{+} \tilde{\nabla}^{2} \rho^{+} -D \tilde\rho^{-} \tilde{\nabla}^{2} \rho^{-}-\\
	&- D \rho^{+}\left(\tilde{\bol{\nabla}} \tilde{\rho}^{+}\right)^{2} - D \rho^{-}\left(\tilde{\bol{\nabla}} \tilde{\rho}^{-}\right)^{2}
\end{split}
\label{mannaggia_bis}
\end{equation}
\begin{equation}
\begin{split}
	\mathcal{S}_{\mathrm{flip}}=&\gamma\, \left(e^{\tilde{\rho}^{+}}-e^{\tilde{\rho}^{-}}\right) \times\\
	&\times\left(e^{-\tilde{\rho}^{+}}\rho^{+}  - e^{-\tilde{\rho}^{-}}\rho^{-} \right)-\\
	+&\frac{\tau}{2}  \left(e^{-\tilde \rho^{+}}-e^{\tilde \rho^{-}}\right) \times\\
	&\times \left(e^{\tilde \rho^{+}} \rho^{+} -  e^{\tilde \rho^{-}} \rho^{+}\right) \rho^{+}\, \rho^{-}\, 
\end{split}
\label{SaAIM_bis}
\end{equation}
We now want to change variables from $\rho^{+}$ and $\rho^{-}$ to $m$ and $\rho$. To do this in the field theory, we must also transform the $\tilde \rho$ fields. It is sufficient for RG purposes to work as usual in a Landau-Ginzburg expansion in fluctuations around the homogeneous disordered state at $m=0,\,\rho=\rho_0$. Hence we shall write $\rho=\rho_0+\delta\rho$, and expand in powers of $m$ and $\delta\rho$.  The resulting action contains an infinite set of nonlinear terms of which only the first few are relevant, in the RG sense, near $4$ dimensions. Retaining only these terms, the result is the sum of a Gaussian action density $\mathcal{S}_{0}$ and a non-Gaussian interaction part $\mathcal{S}_{I}$
\begin{equation}
	\begin{split}
		\mathcal{S}_{0}=
		&
		\tilde m
		\left(\partial_{t}-D\,\nabla^2+a\right)
		m-
		\tlambda\, \tilde m^2 +\\
		&+
		\tilde\rho
		\left(\partial_{t}-D\,\nabla^2\right)
		\delta\rho
		-\tD
		\left(\bol{\nabla}\tilde\rho\right)^{2}
	\end{split}
\label{capitano3gauss}
\end{equation}
\begin{equation}
	\mathcal{S}_{I}=b\,\tilde m\, m^3+g\,\tilde m\, m\,\delta\rho+ {\hbox{\rm irrelevant}}
\label{capitano3int}
\end{equation}
with coefficients derived from microscopic parameters as follows:
\begin{align}
    a&=\frac{1}{4} \left(8 \gamma -\tau \rho_{0}^2  \right)\,, & 
    b&=\frac{\tau}{4}\,, &
    g&=-\frac{\tau\,\rho_{0}}{2}\nonumber\\
    \tlambda&=\frac{\rho_{0}}{4} \left(8 \gamma +\rho_{0}^2 \tau \right)\,, &
    \tD&=\rho_0\,D & & \nonumber
\end{align}
This action can be cast in more familiar form as a pair of Langevin equations, which read
\begin{align}
\partial_{t}m&=D\nabla^{2} m-a\, m-b\, m^{3}-g\,\delta\rho\,m+\sqrt{2\tlambda} \,\eta\label{nonlin1}\\
\partial_{t} \delta\rho &=-\bol{\nabla}\cdot\bol{J}\, ; \quad \bol{J}=-D\,\bol{\nabla} \delta\rho + \sqrt{2\tD}\,\bol{\zeta}\label{nonlin2}
\end{align}
with $\eta$ and $\zeta_{i}$ independent Gaussian white noises of unit variance.

Note that any nonlinearity of the form $\bol{\nabla}(m^2)$ in the current $\bol{J}$ of \eqref{nonlin2}, or equivalently a term $\tilde \rho\, \nabla^2 (m^2)$ in the action \eqref{capitano3int}, if present, would also be relevant in $d<4$. However, since it is absent in the bare theory and there are no other non-Gaussian terms linear in $\tilde \rho$, it will not be generated during an RG transformation. More generally one expects any relevant term, even if absent in the original action, to be generated during the RG flow, unless its absence is protected by some kind of symmetry or conservation law. The physics that prevents the generation of this term in our case is as follows: {\em when $v=0$, the dynamics of the mass density $\rho$ is independent of the state of magnetization $m$}. Such a condition clearly survives coarse-graining, and can arguably be viewed as a symmetry between $A$ and $B$ particles (or up- and down-spins) at microscopic level, stating that the diffusive jump rates of a particle is independent of its spin state. The symmetry is however absent in a model with detailed balance, where the hopping rates {\em must} depend on the energy change caused by the hop, which does depend on the spin state. Since it is possible to construct an AIM that recovers detailed balance at $v=0$ \cite{tal2023}, one cannot view the symmetry found here as fundamental to all AIMs, but it remains a defining feature of all the AIMs studied in this paper  (including AIM0).

\subsection{Connection with Model C}

The stochastic dynamics of Model C are \cite{hohenberg1977theory}
\begin{align}
	\partial_{t}m&=\lambda \nabla^{2} m-\lambda r\, m-\lambda u\, m^{3}-\lambda \gamma \,\delta\rho\,m+\sqrt{2\lambda} \,\eta\nonumber\\
	\partial_{t} \delta \rho &=-\bol{\nabla}\cdot\bol{J}\, \nonumber \\
	 \bol{J}&=-D\,\bol{\nabla} \delta\rho - \frac{D\gamma}{2} \bol{\nabla}(m^2) + \sqrt{2 D}\,\bol{\zeta}\nonumber
\end{align}
Here $\lambda$ is a mobility parameter (unrelated to previous use of the same symbol in this paper), while $r,u,\gamma$ are coefficients in the free energy functional 
$
    \Ham=\int d^dx\,\frac{1}{2}\left(\bol{\nabla} m\right)^2+\frac{r}{2}m^2+\frac{u}{4}m^4+\frac{1}{2}\rho^2+\frac{\gamma}{2}m^2\rho 
$
that underlies the model. Model C obeys detailed balance with respect to this $\Ham$. The noise terms are just as in (\ref{nonlin1},\ref{nonlin2}). 

Strikingly, the {\em only difference} between (\ref{nonlin1},\ref{nonlin2}) for the AIMs under study and Model C is the absence in the AIM case of the term $\bol{\nabla}(m^2)$ in the current $\bol{J}$. As already discussed, this term is relevant but structurally absent in our chosen AIMs, while in contrast it is structurally present, with a coefficient fixed by detailed balance, in Model C. 
The difference between these two cases need not be accessible via any approach that attempts to perturbatively deform one model into the other, for instance by considering small departures from detailed balance. The change in parameters is not small, and moreover replaces one symmetry (time-reversal) with a different and unrelated one (spin-independent density dynamics). 

Interestingly, a generalized model that includes both AIM and Model C  as special cases has previously been introduced and studied using RG methods \cite{akkineni2004nonequilibrium}. The model is defined by
\begin{align}
	\partial_t m&=\lambda \,\nabla^2 m-a\,m-b\,m^3-g_m\,m\,\delta\rho+\sqrt{2\tlambda}\,\eta\label{mModC}\\
	\partial_t \delta\rho&=-\bol{\nabla}\cdot\bol{J}\, \nonumber\\ \bol{J}&=-D\,\bol{\nabla} \delta\rho - \frac{g_\rho}{2} \bol{\nabla}m^2 + \sqrt{2\tD}\,\bol{\zeta}\label{rhoModC}
\end{align}
The AIM2 dynamics of (\ref{nonlin1},\ref{nonlin2}) is recovered as
\begin{align}
	g_\rho&=0 & g_m&=g & \lambda &=D
\end{align}
while equilibrium Model C corresponds to
\begin{align}
	a&=\lambda r & \lambda&=\tlambda & g_m&=\lambda \gamma \label{ceq}\\
	b&=\lambda u & D&=\tD & g_\rho&=D \gamma\label{geq}
\end{align}

\subsubsection{RG flows}

In the present paper we do not review in detail the comprehensive perturbative RG study of this class of models offered by Akkineni and Taueber in \cite{akkineni2004nonequilibrium} (which in fact addresses a much larger class spanning Heisenberg as well as Ising symmetry, and Model D as well as Model C dynamics). 

Briefly, for the model governed by (\ref{mModC},\ref{rhoModC}), various fixed points of potential relevance to AIMs  are considered in \cite{akkineni2004nonequilibrium}. A Gaussian fixed point, stable for $d>4$, becomes unstable for $\epsilon = 4-d>0$. In the absence of $g_m$, the unstable flow is towards a Model A fixed point, at which the $m$ dynamics is decoupled from $\rho$ which is then ignorable. For nonzero $g_m$, however, the Model A fixed point is unstable towards an equilibrium-like Model C fixed point where detailed balance is restored. This is perturbatively stable against detailed-balance violations; its basin of attraction should include all models in which such violation is weak.
Beyond this basin, in addition to the $g_m=0$ manifold where Model A behaviour is recovered, lies a further unstable manifold at $g_{\rho}=0$. The strongly nonequilibrium dynamics on this manifold describes situations, like the AIMs studied here, in which it is the dynamics of $\rho$ that decouples from $m$. On this unstable manifold, a further fixed point was found, whose strongly nonequilibrium dynamics describes a situation in which $m$ relaxes much faster than $\rho$ at large scales. This fixed point is however unstable also within the $g_{\rho}=0$ manifold. Interestingly, Akkineni and Tauber  also found another nonequilibrium fixed point at $g_{\rho}=0$ for which the coupling  $g_m$ seemingly flows to infinity for $d<4$. The latter caused them to conclude that no true nonequilibrium fixed point is accessible at order $\epsilon$ \cite{akkineni2004nonequilibrium}.

Elsewhere \cite{futurepaper}, we calculate the RG flow on the submanifold where $g_\rho = 0$ to which, as we have explained, the AIMs studied in this paper are confined; we argue that despite the conclusions of \cite{akkineni2004nonequilibrium} a nonequilibrium critical point describing the AIM critical point in these strongly nonequilibrium models can be found within perturbative RG approach. 

More importantly for the present discussion, the  $g_\rho = 0$ submanifold does not contain the Model C critical point. This can be seen directly from the following argument. As previously explained, the Model C fixed point splits off from the Gaussian one below $d = 4$. Here the coupling term involving $g_\rho$ is relevant. Only if it were irrelevant could the fixed-point value of this coupling constant become zero at the Model C fixed point. Therefore, this fixed point cannot lie on the $g_\rho = 0$ manifold to which our AIMs are confined. This strongly suggests that, whether or not the AIM critical point is perturbatively accessible to order $\epsilon$ \cite{akkineni2004nonequilibrium,futurepaper}, it should indeed lie in different universality class from Model C. 

This suggestion is different from the one made concerning AIM0 in \cite{solon2015flocking}. The situation is however delicate because, as previously stated, our result depends on a symmetry of all the AIMs considered here (including AIM0 of \cite{solon2015flocking}) which might nonetheless be broken in more general models. Specifically, we know it {\em must} be broken in any AIM that restores detailed balance by construction at the critical point (e.g. \cite{tal2023}), in which case there can be little doubt that the Model C universality class prevails. We also note that numerical evidence favours equilibrium Ising exponents for AIMs in $d=2$ \cite{solon2015flocking} -- which we have also confirmed for ourselves numerically. It is unusual for universality classes to actually merge on reducing dimensionality, so this could indicate that while the Model C and AIM classes retain distinct exponents these are hard to distinguish numerically in two (and therefore possibly three) dimensions.

\section{Conclusion}\label{conc}
We have considered a Doi-Peliti field theoretical formalism, and exploited it to derive an exact field theory able to describe the behaviour of a class of Active Ising Models (AIMs) that allows different choices of the spin-alignment interactions. We showed how field theory provides, as it so often does, a powerful framework to understand collective behaviour in active systems. We were able first to derive several previously known results within this framework. These include the deterministic hydrodynamic equations \cite{kourbane2018exact}; the peculiar behaviour of the two-body collisional interaction, which cannot sustain flocking in the presence of noise \cite{chatterjee2019threebody}; and the linear instability of the homogeneous ordered phase close to the transition, leading to phase-separated profiles and a first order scenario \cite{gregoire2004,chate2008collective,dicarlo2022evidence}. Thereafter we showed how the Doi-Peliti framework can take us far beyond these results. For example, we used it to go beyond the deterministic hydrodynamic equations, complementing them with sub-leading fluctuation terms needed to describe the system on finite scales. Developing the same field theory in a different manner allowed us to address the AIM critical point, defined as the second order alignment transition arising when the self-propulsion term is turned off. We defer to a separate paper a full analysis of the resulting RG flow \cite{futurepaper}. Even without this, we could elucidate the relationship between the critical point of the AIMs studied here and Model C. The latter has the same combination of a nonconserved magnetization with Ising symmetry, coupled to a conserved density, but unlike our AIMs also respects detailed balance. Based on this comparison, we argued that the AIM critical points studied here, contrary to expectation \cite{solon2015flocking}, are {\em not} governed by the the Model C universality class. However, this conclusion stems from a `symmetry' of these particular models whereby diffusive jump rates are not affected by the spin state of a particle. This symmetry need not hold for more general Active Ising Models, and specifically {\em cannot hold} in AIMs constructed so that detailed balance gets restored in the zero self-propulsion limit, such as that of \cite{tal2023}, which can then behave like Model C at criticality.

\backmatter

\bmhead{Acknowledgments}
MEC thanks Fyl Pincus for inspirational discussions on soft matter physics spanning the past 40 years. We thank Rosalba Garcia-Millan for fruitful discussions and Tal Agranov, Robert Jack and Etienne Fodor for a critical reading of the manuscript. MS also thanks Andrea Cavagna and Luca Di Carlo for discussions. This work was funded in part by the European Research Council (ERC) under the EU's Horizon 2020 Programme, Grant agreements No. 740269 and No. 785932.

\onecolumn 
\begin{appendices}

\section{Master Equation}\label{ME}
Here we will give the explicit form of the Master Equation for the different AIMs introduced in Sec. \ref{MastE}.
Since the Master Equation is linear in $P$, each different process gives an independent contribution. For the AIMs considered here the Master Equation can be always written in the form of Equation \eqref{eq:MastE}
\begin{equation}
	\partial_{t}P=\mathcal{L}_{D}\left[P\right]+\mathcal{L}_{\epsilon}\left[P\right]+\mathcal{L}_{\mathrm{flip}}\left[P\right]
\end{equation}
where $\mathcal{L}_{\mathrm{flip}}$ is the contribution coming from spin flipping, while we split in two the contribution of hopping. The first, $\mathcal{L}_{D}$, is obtained by setting $\epsilon=0$ and therefore gives rise to simple, unbiased, diffusion. The second, $\mathcal{L}_{\epsilon}$, is instead the contribution arising from active self-propulsion, and will vanish if $\epsilon$ does. 

\subsection{Hopping}
For all the AIMs introduced here, the dynamics in space is represented by the biased hopping, described by the reactions
\begin{align}
	A_{\bol{i}}&\longrightarrow A_{\bol{i}\pm \hat{x}}\qquad \text{rate: }D(1\pm \epsilon) &
        B_{\bol{i}}&\longrightarrow B_{\bol{i}\pm \hat{x}}\qquad \text{rate: }D(1\mp \epsilon)\\
        A_{\bol{i}}&\longrightarrow A_{\bol{i}\pm \hat{y}} \qquad \text{rate: }D &
        B_{\bol{i}}&\longrightarrow B_{\bol{i}\pm \hat{y}} \qquad \text{rate: }D
\end{align}
These rates for a random jump, expressed in terms of the number of $A_{\bol{i}}$ and $B_{\bol{i}}$ particles, $n_{\bol{i}}^{+}$ and $n_{\bol{i}}^{-}$ respectively, then take the following form
\begin{align}
     A_{\bol{i}}&\to A_{\bol{i}\pm \hat{y}} &
    &W\left(n_{\bol{i}}^{+}-1, n_{\bol{i}\pm \hat{y}}^{+}+1 | n_{\bol{i}}^{+}, n_{\bol{i} \pm \hat{y}}^{+} \right)= n_{\bol{i}}^{+} \, D\qquad\qquad\forall\hat{y}\neq\hat{x}\\
     B_{\bol{i}}&\to B_{\bol{i}\pm \hat{y}} &
    &W\left(n_{\bol{i}}^{-}-1, n_{\bol{i}\pm \hat{y}}^{-}+1 | n_{\bol{i}}^{-}, n_{\bol{i} \pm \hat{y}}^{-} \right)= n_{\bol{i}}^{-} \, D\qquad\qquad\forall\hat{y}\neq\hat{x}\\
     A_{\bol{i}}&\to A_{\bol{i}\pm \hat{x}} &
    &W\left(n_{\bol{i}}^{+}-1,n_{\bol{i}\pm \hat{x}}^{+}+1|n_{\bol{i}}^{+}, n_{\bol{i} \pm \hat{x}}^{+} \right)=n_{\bol{i}}^{+} \, D (1\pm\epsilon)\\
     B_{\bol{i}}&\to B_{\bol{i}\pm \hat{x}} &
    &W\left(n_{\bol{i}}^{-}-1,n_{\bol{i}\pm \hat{x}}^{-}+1|n_{\bol{i}}^{-}, n_{\bol{i} \pm \hat{x}}^{-} \right)=n_{\bol{i}}^{-} \, D (1\mp\epsilon)
\end{align}
Hence, $\mathcal{L}_{D}$ and $\mathcal{L}_{\epsilon}$ always take the form
\begin{equation}
    \begin{split}
        \mathcal{L}_D [P] = D \sum_{\bol{i} } \sum_{\bol{j}:\abs{\bol{i}-\bol{j}}=1} &
    (n_{\bol{i}}^{+}+1)P(\bol{n}^{+}+\bol{1}_{\bol{i}}-\bol{1}_{\bol{j}},\bol{n}^{-};t)
    -n_{\bol{i}}^{+} P(\bol{n}^{+},\bol{n}^{-};t)+\\
    +&(n_{\bol{i}}^{-}+1)P(\bol{n}^{+},\bol{n}^{-}+\bol{1}_{\bol{i}}-\bol{1}_{\bol{j}};t)
    -n_{\bol{i}}^{-} P(\bol{n}^{+},\bol{n}^{-};t)
    \end{split}
\end{equation}
\begin{equation}
\begin{split}
    \mathcal{L}_\epsilon [P]= \epsilon\, D \sum_{\bol{i}} &
    (n_{\bol{i}}^{+}+1)P(\bol{n}^{+}+\bol{1}_{\bol{i}}-\bol{1}_{\bol{i}+\hat{x}},\bol{n}^{-};t)-
    (n_{\bol{i}}^{+}+1)P(\bol{n}^{+}+\bol{1}_{\bol{i}}-\bol{1}_{\bol{i}-\hat{x}},\bol{n}^{-};t)-\\
    -&(n_{\bol{i}}^{-}+1)P(\bol{n}^{+},\bol{n}^{-}+\bol{1}_{\bol{i}}-\bol{1}_{\bol{i}+\hat{x}};t)+
    (n_{\bol{i}}^{-}+1)P(\bol{n}^{+},\bol{n}^{-}+\bol{1}_{\bol{i}}-\bol{1}_{\bol{i}-\hat{x}};t)
\end{split}
\end{equation}

\subsection{Spin-flipping}
\subsubsection{AIM1 rates}\label{aim1}
In the AIM1, the spin-flipping process occurs with rates which are reminiscent of some Ising-like behaviour, namely given by
\begin{align}
        A_{\bol{i}}&\longrightarrow B_{\bol{i}} \qquad
        \text{rate: }\,\gamma\exp{\left(-\beta\, m_{\bol{i}}\right)}=\gamma\exp{\left[\beta\, \left(n_{\bol{i}}^{-} - n_{\bol{i}}^{+}\right)\right]}
        \label{ABceAIM}\\
        B_{\bol{i}}&\longrightarrow A_{\bol{i}} \qquad
        \text{rate: }\,\gamma\exp{\left(\beta\, m_{\bol{i}}\right)}=\gamma\exp{\left[\beta \left(n_{\bol{i}}^{+} - n_{\bol{i}}^{-}\right)\right]}
        \label{BAceAIM}
\end{align}
The global rates associated to this process are
\begin{align}
A_{\bol{i}}&\to B_{\bol{i}} &
    &W\left(n_{\bol{i}}^{+}-1,n_{\bol{i}}^{-}+1|n_{\bol{i}}^{+},n_{\bol{i}}^{-}\right)=
    \gamma\, n_{\bol{i}}^{-}\, \exp{\left(\beta\, (n_{\bol{i}}^{+} - n_{\bol{i}}^{-})\right)}\\
      B_{\bol{i}}&\to A_{\bol{i}} &
     &W\left(n_{\bol{i}}^{+}+1,n_{\bol{i}}^{-}-1|n_{\bol{i}}^{+},n_{\bol{i}}^{-}\right)=
    \gamma\, n_{\bol{i}}^{+}\, \exp{\left(\beta\, (n_{\bol{i}}^{-} - n_{\bol{i}}^{+})\right)}
\end{align}
which contribute to the Master Equation as
\begin{equation}
\begin{split}
    \mathcal{L}_\mathrm{flip}[P]=\gamma \sum_{\bol{i}}\,
    & (n_{\bol{i}}^{+}+1)\exp{\left[\beta \left(n_{\bol{i}}^{-}-n_{\bol{I}}^{+}-2\right)\right]}\, P( \bol{n}^{+} + \bol{1}_{\bol{i}}\,, \bol{n}^{-}-\bol{1}_{\bol{I}}\,;t)-\\
    -&n_{\bol{i}}^{+}\,\exp{\left[\beta\left(n_{\bol{i}}^{-}-n_{\bol{I}}^{+}\right)\right]}\,P(\bol{n}^{+} ,\bol{n}^{-} ;t)+ \\
    + &(n_{\bol{i}}^{-}+1)\exp{\left[\beta \left(n_{\bol{i}}^{+}-n_{\bol{I}}^{-}-2\right)\right]}\,P(\bol{n}^{+} - \bol{1}_{\bol{i}}\,,\bol{n}^{-} + \bol{1}_{\bol{I}}\,;t)-\\
    -&n_{\bol{i}}^{-}\,\exp{\left[\beta\left(n_{\bol{i}}^{+}-n_{\bol{I}}^{-}\right)\right]}\,P(\bol{n}^{+} ,\bol{n}^{-} ;t)
\end{split}
\end{equation}

\subsubsection{AIM2}\label{aim2}
In the case of the AIM2, particles undergo multiple-particle collisions. Here, we consider one-, two- and three-particle collision processes, with rates denoted $\gamma$, $\lambda$ and $\tau$ respectively. These three processes are what we call in the main AIM2.1, AIM2.2 and AIM2.3. In this case, the contribution $\mathcal{L}_{\mathrm{flip}}$ can be further written as
\begin{equation}
\mathcal{L}_{\mathrm{flip}}=\mathcal{L}_\gamma +\mathcal{L}_\lambda+\mathcal{L}_\tau
\end{equation}

{\bf AIM2.1: one-particle collision}\\
The one-particle collision (random) spin-flipping process is defined by the reactions
\begin{align}
        A_{\bol{i}}&\longrightarrow B_{\bol{i}} \qquad
        \text{rate: }\,\gamma &
        B_{\bol{i}}&\longrightarrow A_{\bol{i}} \qquad
        \text{rate: }\,\gamma \ 
\end{align}
The rates for a random spin flipping, expressed in terms of the number of $A_{\bol{i}}$ and $B_{\bol{i}}$ particles, $n_{\bol{i}}^{+}$ and $n_{\bol{i}}^{-}$ respectively, take the following form
\begin{align}
    W(n_{\bol{i}}^{+}+1,n_{\bol{i}}^{-}-1|n_{\bol{i}}^{+},n_{\bol{i}}^{-})&=\gamma\, n_{\bol{i}}^{-}\, \\
    W(n_{\bol{i}}^{+}-1,n_{\bol{i}}^{-}+1|n_{\bol{i}}^{+},n_{\bol{i}}^{-})&=\gamma\, n_{\bol{i}}^{+}\, 
\end{align}
From these rates, we can write down the contribution $\mathcal{L}_{\gamma}\left[P\right]$ of the random spin-flipping to the Master Equation, as
\begin{equation}
\begin{split}
    \mathcal{L}_{\gamma}\left[P\right]=\gamma \sum_{\bol{i}}\,
    & (n_{\bol{i}}^{+} +1)\,P(\bol{n}^{+} + \bol{1}_{\bol{i}}\, , \bol{n}^{-} - \bol{1}_{\bol{i}}\,;t) - n_{\bol{i}}^{+}\,P(\bol{n}^{+},\bol{n}^{-};t)+ \\
    + &(n_{\bol{i}}^{-}+1)\,P(\bol{n}^{+} - \bol{1}_{\bol{i}}\, , \bol{n}^{-} + \bol{1}_{\bol{i}}\,;t) - n_{\bol{i}}^{-}\,P(\bol{n}^{+},\bol{n}^{-};t)
\end{split}
\end{equation}

{\bf AIM2.2: two-particle collision}\\
In terms of reactions, the two-particle collision process can be expressed as
\begin{align}
        A_{\bol{i}}+B_{\bol{i}}&\longrightarrow 2\,B_{\bol{i}} \qquad
        \text{rate: }\,\lambda &
        A_{\bol{i}}+B_{\bol{i}}&\longrightarrow 2\,A_{\bol{i}} \qquad
        \text{rate: }\,\lambda \ 
\end{align}
Note that the two rates must be equal if we want to preserve the symmetry by spin inversion, which is a symmetry under exchange of $A$ and $B$ particles.
The rates associated to this process are
\begin{align}
    W(n_{\bol{i}}^{+}+1,n_{\bol{i}}^{-}-1|n_{\bol{i}}^{+},n_{\bol{i}}^{-})&=\lambda\, n_{\bol{i}}^{+}n_{\bol{i}}^{-}\, \\
    W(n_{\bol{i}}^{+}-1,n_{\bol{i}}^{-}+1|n_{\bol{i}}^{+},n_{\bol{i}}^{-})&=\lambda\, n_{\bol{i}}^{+}n_{\bol{i}}^{-}\, 
\end{align}
which contribute to the Master Equation as
\begin{equation}
\begin{split}
    \mathcal{L}_\lambda [P]=\lambda \sum_{\bol{i}}\,
    & (n_{\bol{i}}^{+} +1)(n_{\bol{i}}^{-} -1)\,P(\bol{n}^{+} + \bol{1}_{\bol{i}}\, , \bol{n}^{-} - \bol{1}_{\bol{i}}\,;t)
    - n_{\bol{i}}^{+} n_{\bol{i}}^{-}\,P(\bol{n}^{+},\bol{n}^{-};t)+ \\
    + &(n_{\bol{i}}^{-}+1)(n_{\bol{i}}^{+} -1)\,P(\bol{n}^{+} - \bol{1}_{\bol{i}}\, , \bol{n}^{-} + \bol{1}_{\bol{i}}\,;t) 
    - n_{\bol{i}}^{+} n_{\bol{i}}^{-}\,P(\bol{n}^{+},\bol{n}^{-};t)
\end{split}
\end{equation}

{\bf AIM2.2: two-particle collision}\\
In terms of reactions, the three-particle collision process can be expressed as
\begin{align}
        2\,A_{\bol{i}}+B_{\bol{i}}&\longrightarrow 3\,A_{\bol{i}} \qquad
        \text{rate: }\,\tau &
        A_{\bol{i}}+2\,B_{\bol{i}}&\longrightarrow 3\,B_{\bol{i}} \qquad
        \text{rate: }\,\tau \ 
\end{align}
Again, if we want the system to obey a symmetry under exchange of $A$ and $B$ species, which is a symmetry under global spin flipping, the rates with which the two reactions take place must be the same.
The global rates associated to this process are
\begin{align}
    W(n_{\bol{i}}^{+}+1,n_{\bol{i}}^{-}-1|n_{\bol{i}}^{+},n_{\bol{i}}^{-})&=\frac{\tau}{2}\, n_{\bol{i}}^{+} \left(n_{\bol{i}}^{+}-1\right) n_{\bol{i}}^{-}\, \\
    W(n_{\bol{i}}^{+}-1,n_{\bol{i}}^{-}+1|n_{\bol{i}}^{+},n_{\bol{i}}^{-})&=\frac{\tau}{2}\, n_{\bol{i}}^{+} n_{\bol{i}}^{-} \left(n_{\bol{i}}^{-}-1\right)\, 
\end{align}
which contribute to the Master Equation as
\begin{equation}
\begin{split}
    \mathcal{L}_\tau [P]=\frac{\tau}{2} \sum_{\bol{i}}\,
    & (n_{\bol{i}}^{+} +1)n_{\bol{i}}^{+}(n_{\bol{i}}^{-} -1)\,P(\bol{n}^{+} + \bol{1}_{\bol{i}}\, , \bol{n}^{-} - \bol{1}_{\bol{i}}\,;t)
    - n_{\bol{i}}^{+} (n_{\bol{i}}^{+} -1) n_{\bol{i}}^{-}\,P(\bol{n}^{+},\bol{n}^{-};t)+ \\
    + &(n_{\bol{i}}^{-}+1) n_{\bol{i}}^{-} (n_{\bol{i}}^{+} -1)\,P(\bol{n}^{+} - \bol{1}_{\bol{i}}\, , \bol{n}^{-} + \bol{1}_{\bol{i}}\,;t) 
    - n_{\bol{i}}^{+} n_{\bol{i}}^{-} (n_{\bol{i}}^{-}-1)\,P(\bol{n}^{+},\bol{n}^{-};t)
\end{split}
\end{equation}

\section{From Master Equation to field theory}\label{DP}
We start by writing the Master Equation for our microscopic model in the form
\begin{equation}
\partial_{t} P(\bol{n},t)=\mathcal{L} \left[P(\bol{n},t)\right] \label{puppets}
\end{equation}
Here  $\mathcal{L}$ is a linear operator acting on $P(\bol{n},t)$ in which $\bol{n}$ specifies a microstate at time $t$. (That is, $\bol{n}$ lists the occupancies of each type of particle at every site in the system.)
The second step is to define a second-quantised Fock space representation: for a single site and particle type we will call $\ket{n}$ the state in which $n$ particles are present. More generally,  we have a Fock state $\ket{n_{1}, n_{2}, n_{3}, \dots}=\ket{\bol{n}}$. The state of the system at time $t$, represented by the probability generating function, can be written as a superposition in Fock space as
\begin{equation}
\ket{\psi(t)}=\sum_{\bol{n}} P(\bol{n},t) \ket{\bol{n}}
\end{equation}
On this Fock space, we can furthermore define a bosonic ladder operator algebra, with a creation operator $a^{\dagger}$ and annihilation operator $a$ that act on the system in the following way
\begin{equation}
a_{i}^{\dagger}\ket{n_{i}}=\ket{n_{i}+1}\qquad a_{i}\ket{n_{i}}=n_{i} \ket{n_{i}-1}
\end{equation}
Note that this normalisation convention differs from that usually introduced in many-body quantum systems. However, the usual commutation relations still hold
\begin{equation}
\Bigl[a_{i},a_{j}\Bigr]=\Bigl[a_{i}^{\dagger},a_{j}^{\dagger}\Bigr]=0\qquad
\Bigl[a_{i},a_{j}^{\dagger}\Bigr]=\delta_{ij}
\end{equation}
In this new notation, we can express $\ket{\psi(t)}$ as
\begin{equation}
\ket{\psi(t)}=\sum_{\bol{n}} P(\bol{n},t) \prod_{i} \left(a_{i}^{\dagger}\right)^{n_{i}} \ket{0}
\end{equation}
where $\ket{0}$ is the vacuum state, where no particles are present.
The evolution of $\ket{\psi(t)}$ is described by an operator $\hat H$ through the relation
\begin{equation}
\partial_{t} \ket{\psi(t)}=-\hat H \ket{\psi(t)}
\label{quantumpuppet}
\end{equation}
where $\hat H$ is the second-quantised version of the operator $\mathcal{L}$ of  \eqref{puppets}. Given $\mathcal{L}$ in \eqref{puppets}, an explicit expression of the operator $\hat H$ in \eqref{quantumpuppet} can be constructed from it.

The action of the Doi-Peliti field theory is then given by
\begin{equation}
    S=\sum_{i} \int dt\, \phi_{i}^*(t)\partial_t \phi_{i}(t) + \int dt \,\frac{\bra{\bol{\phi}^*(t)}\hat{H}\ket{\bol{\phi}(t)}}{\braket{\bol{\phi}^*(t)}{\bol{\phi}(t)}}
    \label{DPaction_app}
\end{equation}
where $\ket{\bol{\phi}}$ and $\bra{\bol{\phi}^*}$ are the coherent states:
\begin{align}
    \ket{\bol{\phi}}&=\left(\prod_{i}e^{\phi_{i} a_{i}^\dagger}\right)\ket{0} &
    \bra{\bol{\phi^*}}&=\bra{0} \left(\prod_{i} e^{\phi_{i}^* a_{i}}\right)
\end{align}
The last step to get an explicit form for the Doi-Peliti action is to compute the second term of \eqref{DPaction_app}, and in particular $\bra{\bol{\phi}^*(t)}\hat{H}\ket{\bol{\phi}(t)}$. This is straightforward if $\hat H$ is normal ordered, but might become a more complicate task when it is not. At a practical level, it is thus usually easier to find the normal ordered representation of $\hat H$ and then substitute $a$ operators with $\phi$ field and $a^{\dagger}$ operators with $\phi^{*}$ fields. If difficulties arise, one can revert to finding the normal ordered form directly by computing $\bra{\bol{\phi}^*(t)}\hat{H}\ket{\bol{\phi}(t)}$, bearing in mind that
\begin{align}
    e^{\phi_{i} a_{i}^\dagger}&=\sum_{l} \frac{\left(\phi_{i}\right)^{l}}{l!} \left(a_{i}^{\dagger}\right)^{n} &
     e^{\phi_{i}^* a_{i}}&=\sum_{l} \frac{\left(\phi_{i}^{*}\right)^{l}}{l!} \left(a_{i}\right)^{n}
\end{align}

\section{Doi-Peliti Action}\label{ciack}
The action of the Doi-Peliti field theory of an AIM will have a form similar to that presented in Appendix \ref{DP}. Since two particle species are present in an AIM, namely $A$ and $B$ particles with $+1$ and $-1$ spin respectively, we will need to take this into account by introducing two sets of creation and annihilation fields. The action therefore reads
\begin{equation}
    S=\sum_{i} \int dt\, \phi_{i}^*(t)\partial_t \phi_{i}(t) +\psi_{i}^*(t)\partial_t \psi_{i}(t) + \int dt \,\frac{\bra{\bol{\phi}^*(t),\bol{\psi}^*(t)}\hat{H}\ket{\bol{\phi}(t),\bol{\psi}(t)}}{\braket{\bol{\phi}^*(t),\bol{\psi}^*(t)}{\bol{\phi}(t),\bol{\psi}(t)}}
    \label{action}
\end{equation}
Where $\hat{H}$ is derived from $\mathcal{L}$ through the procedure described in Appendix \ref{DP}. Since $\mathcal{L}=\mathcal{L}_{D}+\mathcal{L}_{\epsilon}+\mathcal{L}_{\mathrm{flip}}$, therefore also
\begin{equation}
	\hat{H}=\hat{H}_{D}+\hat{H}_{\epsilon}+\hat{H}_{\mathrm{flip}}
\end{equation}
and in turn the action can be written as $S=\sum_{i} \int dt\, \mathcal{S}$ where
\begin{equation}
	\mathcal{S}=\phi_{i}^*(t)\partial_t \phi_{i}(t) +\psi_{i}^*(t)\partial_t \psi_{i}(t) + \mathcal{S}_{D}+\mathcal{S}_{\epsilon}+\mathcal{S}_{\mathrm{flip}}
\end{equation}

\subsection{Hopping}
As done for the Master Equation in Appendix \ref{ME}, we split in two the contribution of hopping also in the field-theoretical action.
\subsubsection{Passive diffusion}
Starting from the diffusive contribution to the Master Equation $\mathcal{L}_D$, we can derive the second-quantised evolution operator associated to it, given by
\begin{equation}
\begin{split}
    \hat{H}_D&=- D \sum_{\bol{i} } \sum_{\bol{j}:\abs{\bol{i}-\bol{j}}=h}
    a_{\bol{j}}^\dagger a_{\bol{i}}- a_{\bol{i}}^\dagger a_{\bol{i}}+b_{\bol{j}}^\dagger b_{\bol{i}}- b_{\bol{i}}^\dagger b_{\bol{i}}=\\
    &=D \sum_{\bol{i} } \sum_{\bol{j}:\abs{\bol{i}-\bol{j}}=h}
    \left(a_{\bol{i}}^\dagger-a_{\bol{j}}^\dagger\right) a_{\bol{i}}+\left(b_{\bol{i}}^\dagger-b_{\bol{j}}^\dagger\right) b_{\bol{i}}=\\
    &=\frac{D}{2} \sum_{\bol{i} } \sum_{\bol{j}:\abs{\bol{i}-\bol{j}}=h}
    \left(a_{\bol{i}}^\dagger-a_{\bol{j}}^\dagger\right) \left(a_{\bol{i}}-a_{\bol{j}}\right)+\left(b_{\bol{i}}^\dagger-b_{\bol{j}}^\dagger\right) \left(b_{\bol{i}}-b_{\bol{j}}\right)\, ,
\end{split}
\end{equation}
Since $\hat{H}_{D}$ is already normal-ordered, the contribution $\mathcal{S}_{D}$ to the Doi-Peliti action density is straightforward to compute, and takes the form
\begin{equation}
   \mathcal{S}_D=
    \frac{D}{2} \sum_{\bol{j}:\abs{\bol{i}-\bol{j}}=h}\,
    \left[\left(\phi_{\bol{i}}^*(t)-\phi_{\bol{j}}^*(t)\right)\left(\phi_{\bol{i}}(t)-\phi_{\bol{j}}(t)\right)+\left(\psi_{\bol{i}}^*(t)-\psi_{\bol{j}}^*(t)\right)\left(\psi_{\bol{i}}(t)-\psi_{\bol{j}}(t)\right)\right]
    \label{DPdiffusion}
\end{equation}
The final contribution to the total action is obtained by summing over sites $\bol{i}$ and integrating over time.

\subsubsection{Active self-propulsion}
Starting from the active self-propulsion contribution to the Master Equation $\mathcal{L}_\epsilon$, we can derive the second-quantised evolution operator associated to it, given by
\begin{equation}
\begin{split}
    \hat{H}_\epsilon&=
    -\epsilon D \sum_{\bol{i} } a_{\bol{i}+h\hat{x}}^\dagger\, a_{\bol{i}} - a_{\bol{i}-h\hat{x}}^\dagger\, a_{\bol{i}} - b_{\bol{i}+h\hat{x}}^\dagger\, b_{\bol{i}} + b_{\bol{i}-h\hat{x}}^\dagger\, b_{\bol{i}}=\\
    &= -\epsilon D \sum_{\bol{i} } \left(a_{\bol{i}+h\hat{x}}^\dagger - a_{\bol{i}-h\hat{x}}^\dagger\right) a_{\bol{i}}-\left(b_{\bol{i}+h\hat{x}}^\dagger -b_{\bol{i}-h\hat{x}}^\dagger\right) b_{\bol{i}}=\\
    &= -\epsilon D \sum_{\bol{i} } \left(a_{\bol{i}+h\hat{x}}^\dagger- a_{\bol{i}-h\hat{x}}^\dagger\right) a_{\bol{i}}-\left(b_{\bol{i}+h\hat{x}}^\dagger -b_{\bol{i}-h\hat{x}}^\dagger\right) b_{\bol{i}}
\end{split}
\end{equation}
Since $\hat{H}_\epsilon$ is already normal-ordered, the contribution $\mathcal{S}_{\epsilon}$ to the Doi-Peliti action density is straightforward to compute, and takes the form
\begin{equation}
     \mathcal{S}_\epsilon=
     -\epsilon\, D\,
     \left[\left(\phi_{\bol{i}+h\hat{x}}^*(t)-\phi_{\bol{i}-h\hat{x}}^*(t)\right)\phi_{\bol{i}}(t)-\left(\psi_{\bol{i}+h\hat{x}}^*(t)+\epsilon\, D\,\psi_{\bol{i}-h\hat{x}}^*(t)\right)\psi_{\bol{i}}(t)\right]
    \label{DPactive}
\end{equation}

\subsection{Spin-flipping}
\subsubsection{AIM1 action}\label{DPaim1}
For the AIM1, we can derive the contribution to the action of spin-flipping starting from $\mathcal{L}_{\mathrm{flip}}$. The second-quantised evolution operator associated to this process is given by
\begin{equation}
    \hat{H}_{\mathrm{flip}}=\gamma \sum_{\bol{i} }\left(a_{\bol{i}}^\dagger-b_{\bol{i}}^\dagger\right)
    \left\{
    a_{\bol{i}}\,
    \exp{\left[\beta\left(b_{\bol{i}}^\dagger b_{\bol{i}} - a_{\bol{i}}^\dagger a_{\bol{i}}\right)\right]}-
    b_{\bol{i}}\,
    \exp{\left[\beta\left(a_{\bol{i}}^\dagger a_{\bol{i}}-b_{\bol{i}}^\dagger b_{\bol{i}}\right)\right]}
    \right\}
\end{equation}
From $\hat{H}_{\mathrm{flip}}$, it is possible to derive the contribution $\mathcal{S}_{\mathrm{flip}}$ to the Doi-Peliti action density by averaging over coherent states, as discussed in Appendix A. This action density takes the form
\begin{equation}
    \mathcal{S}_{\mathrm{flip}}=\gamma\, (\phi^*_{\bol{i}}-\psi^*_{\bol{i}}) 
    e^{-\beta -\psi^*_{\bol{i}}\psi_{\bol{i}} - \phi^*_{\bol{i}} \phi_{\bol{i}}}
    \left(
    \phi_{\bol{i}}\,
    e^{e^{\beta}\psi^*_{\bol{i}}\psi_{\bol{i}} +e^{-\beta}\phi^*_{\bol{i}}\phi_{\bol{i}}}-
    \psi_{\bol{i}}\,
    e^{e^{-\beta}\psi^*_{\bol{i}} \psi_{\bol{i}} + e^{\beta} \phi^*_{\bol{i}}\phi_{\bol{i}}}
    \right)
\end{equation}

\subsubsection{AIM2 action}\label{DPaim2}
In the case of the AIM2, the action $\mathcal{S}_{\mathrm{flip}}$ can be written as
\begin{equation}
\mathcal{S}_{\mathrm{flip}}=\mathcal{S}_\gamma +\mathcal{S}_\lambda+\mathcal{S}_\tau
\end{equation}
We derive in the following each contribution.

{\bf AIM2.1: one-particle collision}\\
Starting from $\mathcal{L}_{\gamma}$, we can derive the associated second-quantised evolution operator \begin{equation}
	\hat{H}_{\gamma}= \gamma \sum_{\bol{i}}\, \left(a_{\bol{i}}^{\dagger}-b_{\bol{i}}^{\dagger}\right) \left(a_{\bol{i}}-b_{\bol{i}}\right)
\end{equation}
which is already normal-ordered. Hence the contribution $\mathcal{S}_{\gamma}$ to the Doi-Peliti action density is straightforward to compute, and takes the form
\begin{equation}
     \mathcal{S}_{\gamma}=  \gamma \, \left(\phi^{*}-\psi^{*}\right) \left(\phi-\psi\right)
    \label{DP1body}
\end{equation}

{\bf AIM2.2: two-particle collision}\\
Starting from $\mathcal{L}_{\lambda}$, we can derive the associated second-quantised evolution operator 
\begin{equation}
	\hat{H}_{\lambda}= -\lambda \sum_{\bol{i}}\, \left(a_{\bol{i}}^{\dagger}-b_{\bol{i}}^{\dagger}\right)^{2}\,a_{\bol{i}}\,b_{\bol{i}}
\end{equation}
Since $\hat{H}_{\lambda}$ is already normal-ordered, the contribution $\mathcal{S}_{\lambda}$ to the Doi-Peliti action density is straightforward to compute, and takes the form
\begin{equation}
    \mathcal{S}_{\lambda}= - \lambda \left(\phi^{*}-\psi^{*}\right)^{2} \phi\,\psi
    \label{DP2body}
\end{equation}

{\bf AIM2.3: three-particle collision}\\
Starting from $\mathcal{L}_{\tau}$, we can derive the associated second-quantised evolution operator 
\begin{equation}
	\hat{H}_{\tau}= -\frac{\tau}{2} \sum_{\bol{i}}\, \left(a_{\bol{i}}^{\dagger}-b_{\bol{i}}^{\dagger}\right)\left[\left(a_{\bol{i}}^{\dagger}\right)^{2} a_{\bol{i}} -\left(b_{\bol{i}}^{\dagger}\right)^{2} b_{\bol{i}}\right] \,a_{\bol{i}}\,b_{\bol{i}}
\end{equation}
Since $\hat{H}_{\tau}$ is already normal-ordered, the contribution $\mathcal{S}_{\tau}$ to the Doi-Peliti action is straightforward to compute, and takes the form
\begin{equation}
    \mathcal{S}_{\tau}= - \frac{\tau}{2} \left(\phi^{*}-\psi^{*}\right)\left[\left(\phi^{*}\right)^{2} \phi - \left(\psi^{*}\right)^{2} \psi \right] \,\phi \,\psi
    \label{DP3body}
\end{equation}

\section{From Langevin equations to field theory}\label{MSR}
The Martin-Siggia-Rose (MSR) formalism \cite{martin1973statistical}, known also as the Janssen-De Dominicis formalism \cite{janssen1976on,de1976techniques}, allows us to describe the behaviour of fields evolving according to stochastic differential equations in terms of a field theory formulated using path integrals.

Let us assume the dynamic behaviour of the field $\vphi$ is defined by the following Ito stochastic differential equation
\begin{equation}
	\bol{\mathcal{F}}\left[\vphi\right]-\bol{\theta}=0
	\label{eq:stochastic_equation}
\end{equation}
with the noise $\bol{\theta}$ characterised by the distribution $P_{\theta}$, while $\vecc{\mathcal{F}}$ is the deterministic evolution operator. The expected value of a given observable $O$ can be computed by averaging $O\left[\vphi\right]$ over the possible noise realisations while requiring $\vphi$ to obey \eqref{eq:stochastic_equation}. This yields
\begin{equation}
	\langle O \rangle= \frac{1}{Z}\int \mathcal{D}\vphi\, O\left[\vphi\right] \int\mathcal{D}\bol{\theta}\, P_{\theta}(\bol{\theta}) \, \delta(\bol{\mathcal{F}}\left[\vphi\right]-\bol{\theta})
\end{equation}
where
\begin{equation}
	Z=\int \mathcal{D}\vphi \int\mathcal{D}\bol{\theta}\, P_{\theta}(\bol{\theta}) \, \delta(\bol{\mathcal{F}}\left[\vphi\right]-\bol{\theta})\, 
\end{equation}
By means of an integral representation of the delta-function $\delta(x)=\frac{1}{2\pi}\int \di\tilde{x}\, e^{\iu\,\tilde{x}\, x}$, we can write $\langle O \rangle$ as
\begin{equation}
	\langle O \rangle = \frac{1}{Z} \int \mathcal{D}\vphi \int \mathcal{D}\hvphi\, O\left[\vphi\right] \,e^{ -\iu \hvphi \cdot\bol{\mathcal{F}}\left[\vphi\right]}  \langle e^{\iu \hvphi\cdot\bol{\theta}}\rangle_{\theta}=
	 \frac{1}{\iu Z}\int \mathcal{D}\vphi \int \mathcal{D}\hvphi\, O\left[\vphi\right] \,e^{ - \hvphi \cdot\bol{\mathcal{F}}\left[\vphi\right]+K_{\theta}[\hvphi]}
\end{equation}
where $\langle \,\cdot\, \rangle_{\theta}=\int\mathcal{D}\bol{\theta}\,\cdot\, P_{\theta}(\bol{\theta}) \,$ in the average over noise realisations. In the second equality we performed the substitution $\hphi\to \iu\hphi$. Moreover, $K_{\theta}[\bol{x}]=  \ln\left[\langle e^{\iu \hvphi\cdot\bol{\theta}}\rangle_{\theta}\right]$ is the cumulant-generating function of the distribution $P_{\theta}$. In the case of a Gaussian distribution with zero mean and covariance matrix $2 L_{\alpha\beta}$, the cumulant-generating function is given by $K_\theta[\bol{x}]=x_\alpha L_{\alpha\beta}x_\beta$.

The outcome of this algebraic manipulation is that the statistics generated by the dynamical behaviour defined through \eqref{eq:stochastic_equation} is reproduced by the field-theoretical action $S$ given by,
\begin{equation}
	S[\vphi,\hvphi]=\int\di\vx\; \di t\; \hvphi\cdot \bol{\mathcal{F}}\left[\vphi\right] - K_\theta[\hvphi]
	\label{eq:eff_action}
\end{equation}
The introduction of the auxiliary field $\hvphi$ in the action is the price that has to be paid to exploit the path integral formulation, using the standard rules of static renormalisation and writing the perturbative series in terms of Feynman diagrams.
The field theoretical description reproduces the stochastic dynamics in the sense that, for a given observable $ O\left[\vphi\right]$,
\begin{equation}
	\langle  O \rangle=\langle O \rangle_{S}
\end{equation}
where $\langle O \rangle$ is the average value of $ O$ over all possible realisations of the noise $\bol{\theta}$, while 
\begin{equation}
	\langle O \rangle_{S}=\frac{1}{\mathcal Z} \int\mathcal{D}\vphi\int\mathcal{D}\hvphi\, O\left[\vphi\right]
	\eu^{-S[\vphi,\hvphi]}
\end{equation}
is the average over the field-theoretic action.
Thanks to this equivalence, the dynamics can be investigated by studying the action $S$ through field-theoretical techniques, including perturbation theory and the renormalisation group.

The Gaussian part of the action $S$ corresponds to the linear dynamics, namely the linear part of the operator $\mathcal{F}$, while the interactions derive from non-linear terms. Within this formalism, an external source $\vecc{h}$ introduced in the dynamical equation of $\vphi$ is coupled to $\hvphi$ in the effective action. Therefore the response function can be written as
\begin{equation}
	\dder{\langle\phi_\alpha\left(\vx,t\right)\rangle}{h_\beta\left(\vx',t'\right)}=\langle\phi_\alpha\bigl(\vx,t\bigr)\tilde{\phi}_\beta\bigl(\vx',t'\bigr)\rangle
	\label{eq:response}
\end{equation}
For this reason, $\hvphi$ takes the name `response field'.

Note, crucially, that the equivalence derived in this section works also the other way round: if one obtains a field theoretical action like \eqref{eq:eff_action} from other approaches, say a Doi-Peliti approach, the resulting behaviour of $\vphi$ is equivalent to that which would arise, if $\vphi$ obeyed \eqref{eq:stochastic_equation}. In general, the Doi-Peliti action allows for non-Gaussian noises such as the Poisson noise linked to discrete chemical reactions or other jumps; these are represented in the Master Equation but cannot be written in strict Langevin equation form for which noise is generally assumed to be Gaussian and white. (Note that \eqref{eq:stochastic_equation} is more general than this.) Nonetheless, it is common to recover a standard Langevin equation with Gaussian noise in particular limits, such as the high-density limit, or indeed the hydrodynamic limit as derived in Sec \ref{flu}.

\section{Continuum Limit}\label{cont}%
We will here develop the continuum limit of our theory. To do so, we will need to send the lattice spacing $h\to 0$, paying attention to the most physical way to rescale the various parameters in this limit. Let us first change the spatial variable used to describe the system. Instead of using the lattice number $\bol{i}$, we will use the continuous variable $\bol{x}=h\bol{i}$. Sums over lattice sites $\bol{i}$ are thus replaced by integrals over $\bol{x}$, with the prescription
\begin{equation}
	\int d\bol{x} f(\bol{x}) = h^{d} \sum_{\bol{i}} f_{\bol{i}}\, , \qquad f(h\bol{i})=f_{\bol{i}}\, .
\end{equation}
Second, we must carefully define the continuous-space fields. Say that we have already performed a Cole-Hopf transformation, and have a theory that described the behaviour of the number of particle field $\rho$. In case of more particle species, as it is for AIMs, the same procedure described here can be easily generalised.

While we leave the tilde fields $\tilde\rho$ unchanged, it is convenient to rescale the $\rho$ field with the volume around each lattice site $h^{d}$. This comes from the fact that, when the $h\to0$ limit is taken at fixed total number of particles, the expected number of particles on each site vanishes. What remains constant is the density, namely the average number of particles divided by the volume occupied by a single site $h^{d}$. Hence, it is convenient to rescale the annihilation fields by $h^{d}$, namely
\begin{align}
	\tilde\rho(h\bol{i},t)&=\tilde\rho_{\bol{i}}(t) & \rho(h\bol{i},t)&=h^{-d}\rho_{\bol{i}}(t)
\end{align}
Moreover, by observing that that 
\begin{equation}
	f(\bol{x})-f(\bol{x}+h\hat{y})=-h\, \partial_{\hat{y}}f(\bol{x})+o(h)\, ,
\end{equation}
it is straightforward to write the action contributions that describing processes involving different sites, namely the hopping process in case of AIMs.

The final choice  to be made concerns the scaling with $h$ of the reaction and jump rates. For example, in order to keep finite the mean square displacement due to the diffusive process,  the microscopic hopping rate $D$ must diverge as $h^{-2}$. Similarly, to make the propulsion speed not diverge in the continuum limit, we require weak microscopic bias, namely that $\epsilon$ vanishes with $h$. For the case of multiple-body interactions, all the rates must be rescaled in order to compensate the vanishing probability of finding more than one particle on the same site. For example, the rate of an $n$-body interaction should diverge as $ h^{-d(n-1)}$ in order to guarantee that a spin-flipping due to the $n$-body interaction occurs on average on the same time-scale as the other processes.

\end{appendices}

\end{document}